\documentclass[preprint2]{aastex}
\usepackage{graphicx}
\usepackage{natbib}
\bibpunct{(}{)}{;}{a}{}{,}
\shorttitle{IC~1795}
\shortauthors{Roccatagliata V. et al.}

\begin{document}

\title{Disk Evolution in OB Associations -\\ Deep {\it Spitzer}/IRAC Observations of
  IC~1795} 


\author{Veronica Roccatagliata \altaffilmark{1,2}, Jeroen
  Bouwman\altaffilmark{1}, Thomas Henning\altaffilmark{1}, Mario
  Gennaro\altaffilmark{1}, 
  %Leisa Townsley\altaffilmark{2}, 
  Eric Feigelson\altaffilmark{3,4}, Jinyoung Serena Kim\altaffilmark{5}, Aurora Sicilia-Aguilar\altaffilmark{1}, Warrick
  A. Lawson\altaffilmark{6}%, Hendrik Linz\altaffilmark{1}, Konstantin
  %V. Getman\altaffilmark{2}, Kevin Luhman\altaffilmark{2,3}, Junfeng Wang\altaffilmark{6}
  } 
%\affil{Max-Planck-Institut f\"ur Astronomie (MPIA), 
%              K\"onigstuhl 17, D-69117 Heidelberg, Germany}
%\email{roccata@mpia.de}

%\author{...\altaffilmark{4,5}}
%\affil{...}
%\email{}

%\and

%\author{...h\altaffilmark{5}}
%\affil{}


\altaffiltext{1}{Max-Planck-Institut f\"ur Astronomie, 
              K\"onigstuhl 17, D-69117 Heidelberg, Germany}
  \altaffiltext{2}{Space Telescope Science Institute, Baltimore, MD 21218, USA}            
\altaffiltext{3}{Department of Astronomy and Astrophysics, Pennsylvania State
  University, University Park, PA 16802, USA} 
\altaffiltext{4}{Center for Exoplanets and Habitable Worlds, 525 Davey Laboratory,
 Pennsylvania State University, University Park, PA 16802, USA}
\altaffiltext{5}{Steward Observatory, University of Arizona, 
933 N. Cherry Ave. Tucson, AZ 85721-0065
}
\altaffiltext{6}{School of Physical, Environmental, and Mathematical Sciences,
  University of New South Wales, Australian Defence Force Academy, Canberra,
  ACT 2600, Australia}
%\altaffiltext{6}{Harvard-Smithsonian Center for Astrophysics, 60 Garden
%  Street, Cambridge, MA 02138, USA } 
%\altaffiltext{5}{...}


\begin{abstract}

We present a deep {\it Spitzer}/IRAC survey of the OB association IC~1795
carried out to investigate the evolution of protoplanetary disks in regions
of massive star formation.\\ 
Combining {\it Spitzer}/IRAC data with Chandra/ACIS observations, we find 289
cluster members. An additional 340 sources with an infrared excess, but without X-ray
counterpart, are classified as cluster member candidates. Both surveys are
complete down to stellar masses of about 1~M$_\odot$. 
We present pre-main sequence isochrones computed for the first time in the
{\it Spitzer}/IRAC colors. The age of the
cluster, determined via the location of the Class III sources in the
[3.6]-[4.5]/[3.6] color-magnitude diagram, is in the range 
of 3 - 5~Myr.  \\
%ship is accurately determined by matching infrared and
%X-ray observations with {\it Spitzer}/IRAC and Chandra/ACIS respectively.\\
%IRAC sources with infrared excess (characteristic of circumstellar disk
%around young star) without X-ray counterpart are classified as cluster
%member candidates.\\
%The spatial distribution of the cluster was found to be not symmetric
%around the cluster center, nor around the massive O and B type stars.\\
As theoretically expected, we do not find any systematic variation in
the spatial distribution of disks within 0.6 pc of either O-type star in the association.
%within 0.6~pc from the O6.5V star in the association. 
 However, the disk fraction in IC~1795 does depend on the stellar mass: 
 sources with masses $>$2~M$_\odot$ have a disk fraction of $\sim$20\%, while lower mass
objects (2-0.8~M$_\odot$) have a disk fraction of $\sim$50\%. This implies that disks around massive stars
 have a shorter dissipation timescale. 
\end{abstract}

\keywords{ISM: individual (W3); open clusters and associations: individual (W3
  IC1795); planetary systems: protoplanetary disks; stars: formation; stars:
  pre-main sequence; X-rays: stars} 

\section{Introduction}
\label{intro}
%Circumstellar (CS) disks, surrounding T Tauri and Herbig Ae/Be
%pre-main sequence (PMS) stars, are believed to be the sites of ongoing planet
%formation. These protoplanetary disks have  
%masses and sizes comparable to the expected values for the primitive
%solar nebula \citep[e.g.][]{Beckwithetal1990, AndrewsWilliams2005}. 
Circumstellar (CS) disks have been found around all types of objects, ranging from Herbig Ae/Be
stars to brown dwarfs, sharing similar dust composition, although differences between the disks of low-mass stars / brown dwarfs, solar type stars and more massive stars
seem to exist in the evolution of both the dust and gas components \citep[e.g.][]{HenningMeeus2009, Pascuccietal2009}.
Though the time-scales of disk evolution and dispersion are not well constrained,
90\% of the PMS systems appear to disperse the dust and gas within the planet formation zone
 within $\sim$10 Myr
\citep[e.g.][]{Mamajeketal2004,Bouwmanetal2006,Fedeleetal2010}. Consequently, the formation of gas planets has to be completed within this time-scale.  \\
Although most solar-type and low-mass stars form in OB associations,
disks studies around PMS stars deal mostly with regions of low-mass star formation in
the Gould Belt \citep[e.g.][]{hartmannetal2005, Winstonetal2007,
  Hatchelletal2007, Sicilia-Aguilaretal2008, Evansetal2009} and
nearby moving groups \citep[e.g.][]{Lyoetal2003, Sicilia-Aguilaretal2009, Meeusetal2009}. \\
An exception is the Orion Nebular Cluster \citep[e.g.][]{Hillenbrand1997, Hillenbrandetal1998}, the nearest \citep[$\sim$410~pc;][]{Jeffries2007}, young, %\citep[$\approx 1$ Myr;][]{Hillenbrand1997},
dense and rich cluster with the O6 star, $\theta^1$Ori\,C, in the center. The disks are seen as dark silhouettes against the bright nebula background and they are called ``proplyds'' \citep[e.g.][]{OdellWong1996}. 
Studies of the disk fraction in Orion \citep[e.g.][]{Hillenbrandetal1998} reveal a higher number of disks within 0.5 pc than at larger distances from the O6 star.  
This suggests that the central O star has had insufficient time to fully photoevaporate surrounding disks. 
However,  \cite{Eisneretal2008} found that the percentage of disks more massive than $\sim$0.01 M$_{\odot}$ is lower than in Taurus. \\
There is compelling observational evidence that these disks have been influenced by the ultraviolet radiation from  $\theta^1$Ori\,C \citep[e.g.][]{Johnstoneetal1998, Clarke2007, Gortietal2009}, in that  
the ionized gas surrounding these disks has a cometary-shape \citep[][]{McCulloughetal1995, Ballyetal1998}, their mass distribution is truncated at 0.03~M$_\odot$ and the outer radii, in most of the cases, are smaller  than 60~AU % and the highest value was found to be 150~AU, in correspondence with the most massive disk of $\sim$0.03~M$_\odot$ 
\citep[][]{MannWilliams2009a}. 
However, such a truncation in mass and radii distribution might be due to disk mass loss triggered by stellar encounters which are also responsible for the destruction of 10-15\%  disks in the Trapezium cluster \citep[e.g][]{ScallyClarke2001, Olczaketal2006}. 
\\
%Close to the Trapezium stars direct evidence for the photoevaporation of disks has been found (?? REF!).\\
%Evidence for photoevaporation of disks has also been found by
%\cite{Stecklumetal1998}. They found that G5.97-1.17, a
%young star surrounded by a circumstellar disk, is being photoevaporated by
%the O7 star, Herschel 36, in the center of the Lagoon Nebula.\\
Investigations of disk dissipation processes induced by massive stars, such as photoevaporation and dynamical destruction, must be made in rich, crowded young clusters.
To properly carry out such a study, a disk-unbiased census of cluster members and a robust tracer of disks are needed. 
Chandra X-ray surveys provide member lists based on pre-main sequence magnetic activity; these are particularly effective in locating the disk-free (Class III) stellar population \citep{Feigelsonetal2007}. Spitzer mid-infrared surveys provide sensitive member lists based on infrared excess; these are particularly effective in locating the disky (Class 0-I-II) stellar population. The X-ray and infrared approaches are complementary in other respects: the X-ray selected samples nicely discriminate young cluster members from older Galactic field stars that dominate infrared surveys, while the infrared samples detect the very low-mass population with magnetic activity too faint to be detected in most X-ray exposures.

%To properly carry out such a study, a confirmed cluster membership is needed and a robust disk emission tracer is required.\\
The characterization of circumstellar material in the inner part of the disks can be provided by sensitive IRAC surveys, while a complete membership may be identified by an optical spectroscopy survey for low-mass members with spectral types K -- M \citep[via the Li emission at 6708\AA, e.g. ][]{Jeffriesetal2007, Mentuchetal2008} or accreting objects \citep[via H$_\alpha$ and UV-excess emission, ][]{Fedeleetal2010, Sicilia-Aguilaretal2010, Fangetal2009}. The solar-type mass population can be identified via an X-ray survey.\\
\subsection{Past work}
This approach has be used by, e.g., \citet[][]{Damianietal2006}, who combined a deep X-ray Chandra survey with an optical spectroscopic survey to study the population properties of the 5~Myr old cluster NGC~2363, which hosts an O9.5I star. They found that only 5-9\% of the cluster members are still accreting.  \\
%Optical spectroscopic surveys are sensitive only to accretors, yet not all stars with disks are accreting \citep[e.g.][]{Sicilia-Aguilaretal2010, Fedeleetal2010}. Information to complete the disk census of a cluster is provided by infrared surveys searching for infrared excess emission attributable to protoplanetary disks.\\
%Since not all the stars with disks are accreting into the central star \citep[e.g.][]{Sicilia-Aguilaretal2010, Fedeleetal2010}, a complementary information about the disk fraction is provided by infrared surveys.\\
Recently \cite{Balogetal2007} studied disk evolution and the influence of the O stars in the high-mass star-forming region NGC~2244. Using IRAC observations, they found an overall disk fraction of 44.5\%, but that the fraction of sources with disks was lower ($\sim 27$ \%) within 0.5 pc of the O stars.\\
\cite{Guarcelloetal2009, Guarcelloetal2007} analyzed the disk fraction in the high-mass OB association NGC~6611, which hosts 56 stars with spectral type earlier than B5. The cluster membership and the disk-fraction were defined combining optical, near- and mid-infrared and X-ray studies. The disk fraction for the O and B stars was found varying between 31\% and 16\% across a complete range of incident fluxes up to 800~$erg\,cm^{-2} s^{-1}$. 
In a smaller region compared to the \citet{Guarcelloetal2009, Guarcelloetal2007} studies, \citet{Oliveiraetal2005} computed a much higher disk fraction ($\sim$58\%) probably due to the lower sensitivity of their survey.\\
\citet{Getmanetal2009} studied a portion of the Cep OB3b cluster and the triggered population in the adjacent Cep B molecular cloud. Using a combined $Chandra - Spitzer$ sample, they found that the disk fraction showed a spatial gradient from 75\% in the cloud core to 25\% in the unobscured cluster. They attributed the gradient to stellar ages rather than photoevaporation.\\
%, as no UV-luminous O stars lie in the immediate vicinity.\\
\citet{Hernandezetal2008} computed the disk fraction in the 5~Myr old $\gamma$Velorum cluster, with a central binary consisting of an O7.5 star and a Wolf-Rayet star. Only 5\% of cluster members showed an infrared excess.  \\
%The effect of the photoevaporation is mostly limited to the immediate vicinity
%of the massive stars. Calculations by \citet{Johnstoneetal1998} show that the
%``sphere of influence'' is not much larger than 0.3~pc around an O6
%star. 
%Models of \cite{RichlingYorke2000, Adamsetal2004, Matsuyamaetal2003} showed evaporation acting in
%different regions of the disk: According to different simulations that span between 50-100~AU and  5-10 AU and about 2 AU.
Despite these studies, the number of investigations concerning the presence of disks around stars in OB associations has been limited.\\

\subsection{This work}
In this paper, we present the analysis of protoplanetary disks around young
stars in the IC~1795 OB association. We obtained infrared imaging data of the
IC~1795 OB association with the Infrared Array Camera \citep[
    IRAC;][]{Fazioetal2004} on board the {\it Spitzer Space Telescope}
\citep[][]{Werneretal2004}. To define the cluster membership of IC~1795 we
carried out a deep survey with the Advanced CCD Imaging Spectrometer (ACIS)
detector on board the {\it Chandra X-ray Observatory} \citep{Weisskopfetal2002}.\\  
The IC 1795 cluster ionizes the diffuse HII region in the W3 giant
molecular cloud complex. This region is located in the Perseus arm, which contains several 
spectacular regions of high- and low-mass star formation: W3-North, W3-Main
and W3-OH. 
IC~1795 hosts a bright O6.5V star, BD~+61$^\circ$411, originally spectrally classified by \citet{Mathys1989}, a O9.7I  star and two B stars which have been spectrally classified in the optical by \citet{Oeyetal2005}.
The cluster is assumed to lie at the same distance as W3-OH. From maser 
kinematics this has been accurately measured to be 
2.0 kpc \citep{Xuetal2006,Hachisukaetal2006}. \\
\cite{Oeyetal2005}
derived an approximate age of 3 - 5~Myr and they propose that
this OB association, triggered by the neighboring W4 superbubble, is triggering
new star formation in the young massive regions W3-North, W3-Main and W3-OH. \\
The X-rays morphology of these regions suggests, however, that only the W3-OH
structures are consistent with the collect-and-collapse triggering process 
caused by shocks from the older IC 1795 cluster \citep{FeigelsonTownsley2008}. 
In the X-ray maps, the embedded W3-Main cluster does not show the elongated 
and patchy structure of a recently triggered star cluster, and instead it appears to 
have formed in an earlier episode. \\
  Previous {\it IRAC} observations of the entire W3 region, 
  presented by \cite{Ruchetal2007}, revealed that a large fraction
  of Class~II sources lie within the central cluster IC~1795. No analysis
  of the spatial distribution and disk fraction of the cluster has been performed
  so far. The photometric observations presented in this paper concentrate on the lightly obscured 
  IC~1795 OB association, and 
  they have one order of magnitude better sensitivity 
  %are one order of magnitude deeper 
  compared with the \cite{Ruchetal2007} observations. A new deeper X-ray {\it Chandra} survey of IC~1795 is also presented in this work to accurately define the cluster membership.\\
  \noindent
This paper is organized as follows: in section~\ref{obs} we present our {\it
  Spitzer} and {\it Chandra} observations, along with the data
reduction. 
Results are shown in section~\ref{results}. 
Sections~\ref{membership} and \ref{analysis} include the analysis of the cluster 
(membership, age, infrared properties of the stars).
In section~\ref{discussion} we discuss our results in terms of disk evolution.
Conclusions are drawn in section~\ref{conclusions}. 

\section{Observations and data reduction}
\label{obs}
%IC~1795 was observed with {\it Spitzer}\footnote{{\it Spitzer} program
%  ID~30726, PI J. Bouwman} and {\it Chandra}\footnote{{\it Chandra} program
%  ID~7356, PI J. Bouwman}.  
\subsection{{\it IRAC} observations}
\label{iracobs}
IC~1795 was observed in 2007 September \footnote{{\it Spitzer} program
 ID~30726, PI J. Bouwman} with all 4 channels
centered at 3.6~$\mu$m, 4.5~$\mu$m, 5.8~$\mu$m, 8.0~$\mu$m respectively, using
a 3x4 mosaic pattern, with pointings separated by 220\arcsec\, and aligned
with the array axes. 
The resulting images provide full coverage over a 14\arcmin\,x12\arcmin\,area in all four IRAC channels.\\  
The images have been obtained in High Dynamical Range Mode
in order to obtain unsaturated measurements for all observed cluster members.
The maps are obtained in 144 cycles organized in 12-point dither patterns.
In each channel, a short exposure of 144 frames of 0.4~$s$ each
and a long exposure of 10.4~$s$ per 144 frames have been carried out.  
In each channel, the
average short and long exposures over the entire mosaic are 58 and 1497 sec, respectively. 
The long and the short exposures are analyzed separately to avoid 
saturation of bright sources in the long exposures. \\
The raw data were processed and calibrated with the IRAC pipeline (version
S16.1.0) and the Basic Calibrated Data (BCD) were downloaded from the {\it Spitzer}
archive\footnote{http://archive.spitzer.caltech.edu/}.  
The final mosaics were created using the MOPEX pipeline (version
18.2.2)\footnote{http://ssc.spitzer.caltech.edu/postbcd/download-mopex.html}:
the software takes the individual BCD frames and combines them to
create a mosaic of the observed region, removing cosmic rays and bad pixels
from the single frames.  
%In Fig.~\ref{ch1ch2}-\ref{ch3ch4} 
The combined mosaic of the long exposures at
3.6~$\mu$m/4.5~$\mu$m/8.0~$\mu$m of IC~1795 
is shown in Fig.~\ref{prova}, together with the positions
of the center and the edge of the cluster. %, the high-mass 
%stars spectrally classified \citep[][]{Oeyetal2005}.
%combined regions with the center
%and the edge of the cluster (discussed in section \ref{IRACpositions}). 
 Due to the mosaic configuration, the exposures of ch1 and ch3 cover
 part of W3-Main, while ch2 and ch4 cover part of W3-OH.

\subsection{{\it IRAC} photometry}
\label{iracphot}
We used the APEX/MOPEX package
%to determine the ``point-spread function'' 
to perform PSF-photometry for every detection. 
In particular, we used the APEX MultiFrame pipeline, where the detection of
the point sources is done on the final mosaic in order to recover also the 
fainter objects, while the PSF photometry is carried out on each frame
separately. \\
Over the cluster region, the background was highly variable. 
We estimated the background variability by considering six different regions of the cluster of 25x25 pixels each, corresponding to an area of $\sim$30\arcsec\,x\,30\arcsec. 
%The estimate of such variability was done on six different regions of the cluster on a surface of 25x25 pixels each. 
The mean background flux at 3.6~$\mu$m was about 0.07~Jy with an average standard deviation of 0.05~Jy. 
%These values increase with wavelengths up to 1.4~Jy and a standard deviation of 0.2~Jy at 8~$\mu$m. 
At the edge of IC~1795, where it borders on the younger regions W3-OH on the south-west direction and on W3-Main on the North-East direction, the background increases up to 0.08~Jy and 1.6~Jy at 3.6 and 8~$\mu$m, respectively. Here the background is most likely associated to the molecular clouds of the younger W3-OH and W3-Main regions. Over the cluster region it is instead associated with the PAH emission which peaks at 3.3, 6.3, 7.7 and 8.6~$\mu$m.\\
 As the background was highly variable, PSF photometry was computed in a
small area of 8x8 pixels %(default value is 25x25) 
and subtracted before performing PSF-photometry.
Instead of PSF fitting, APEX uses the point-response function (PRF)
fitting method. The PRF is the response of the detector array pixels to a point
source and combines information on the PSF, the detector sampling and the
intra-pixel sensitivity variation. We used the standard PRFs provided in
APEX. 
The point-source positional uncertainties and the flux uncertainty ({\it
  $\Delta_{PRF}$}) are computed from the covariance matrix calculated from the
``best-fit" to the data. \\
The calibration uncertainties \citep{Reachetal2005} are dominated by the array-location-dependent
photometric correction and the pixel phase effect which might introduce an uncertainty of up to 10\% of the
flux. However, since the pixel phase effect decreases as the
square root of the number of dithers (12 in our case), we conservatively adopt
a value of 5\%.  
The color correction and the absolute calibration error are of on the order of a few \%. 
Overall the final flux calibration error ({\it $\Delta_{cal}$}) is smaller
than 10\% of the flux.  
The total flux uncertainty
is:\\ $\Delta_{flux}=\sqrt{\Delta_{PRF}^2+\Delta_{cal}^2}$.\\ 
Only sources with a signal-to-noise (S/N) flux measurement greater than 6 in each IRAC 
channel have been considered in our final catalog. The final photometric catalog was created, 
combining the results of the analysis of the short and long exposures separately. For objects 
detected in both exposures (and not saturated), we adopt the photometry measured in the long exposure. 
The detection limit of our IRAC survey in the 3.6~$\mu$m band is
16.5~mag. However, we have cases where the high variability of the background
did not enable us to detect all the sources with brightness between
15.5-16.5~mag at 3.6~$\mu$m. 
%The {\it Spitzer}/IRAC catalog is hence $\sim$90\% complete down to 15.5~mag at 3.6~$\mu$m. 
%At longer wavelengths the $\sim$70\% completeness level at
At 4.5~$\micron$ we detected sources down to 16~mag, at 5.8~$\micron$ 13.3~mag, and at 8.0~$\micron$ 11.5~mag.  

\subsection{{\it ACIS} observations}
\label{X-rayobservations}

Our X-ray catalog was generated from a mosaic of {\it Chandra} observations
consisting of the 7 exposures of W3 presented by Feigelson \& Townsley (2008)
and an additional exposure of IC~1795.  This is a 50.0~ks exposure obtained on
2007 December 4 centered on $(\alpha_{J2000}, \delta_{J2000})$ =
(02h:26m:33.6s,+62$^{o}$:00$\arcmin$:35.9$\arcsec$) with the 4-CCD ACIS Imaging Array subtending $17\arcmin
\times 17\arcmin$. The data were reduced using procedures implemented in the
IDL-based {\it Tools for ACIS Review and Analysis (TARA)} and {\it ACIS Extract}
software packages.  The algorithms are described in \citet{Broosetal2010} and are
available online\footnote{http://www.astro.psu.edu/xray/docs/TARA}.  Their
application to stellar populations in other star-forming regions include M~17
\citep[][]{Broosetal2007}, the Rosette Nebula \citep{Wangetal2008,
  Wangetal2009}, and NGC~6334 \citep{Feigelsonetal2009}.      \\
Summarizing briefly here, the {\it Chandra} satellite telemetry data are
subject to various cleaning operations, and the image is aligned to the
2MASS/{\it Hipparcos} astrometric frame.  A superset of tentative sources is
identified by a combination of wavelet-based source detection, peaks in image
tiles with point spread function removed by a maximum-likelihood image
reconstruction, and visual inspection for close multiple sources.   Photons
are extracted from each tentative source in the broad soft, hard and total
X-ray bands, and photon counts are compared to the locally measured background.
Local backgrounds are essential due to the different exposures from
overlapping ACIS exposures.  Source significance is evaluated statistically
 by the quantity $P_B$, the probability based on Poisson statistics that the
 observed source counts would be present given the observed local background
 rate.    \\
Choosing a source significance limit $P_B < 1$\% in at least one broad band,
we find 2192 X-ray sources in the {\it Chandra} mosaic covering the W3-North,
W3-Main, W3-OH, and IC~1795 regions.  Source properties including X-ray
spectra, absorptions, luminosities, and variability, will be presented in a
forthcoming  paper (Townsley et al., in preparation).  For the present study, we
consider only the X-ray source positions and error circles.  Typical estimated
$1\sigma$ error circle radii are $0.2\arcsec\,-\,0.5$\arcsec\, but can exceed 1\arcsec\/
for sources far off-axis.  Note that the {\it Chandra} positional errors are
often considerably smaller than IRAC positional errors shown in Fig.~\ref{coof}
(panels A and B).   
\subsection{Near-infrared survey of IC~1795}
\label{Near-infraredandoptical}
 %Optical and 
Near-infrared photometry of IC~1795 has been also analyzed. %Optical {\it V}-
%and {\it I}-band photometry in the Bessell system were obtained with the
%90Prime imager \citep{Williamsetal2004} at the Bok 2.3m telescope at Steward
%magnitude is 22~mag in {\it V}. Detailed informations about observations and
%data reduction will be described in a forthcoming paper by Jose' et al. (in
%preparation).\\
  Near-infrared photometry in {\it J} (1.25$\mu$m),
{\it H} (1.65$\mu$m) and {\it Ks} (2.17$\mu$m) bands is taken from the Two
Micron All Sky Survey \citep[2MASS][]{struskie2006}\footnote{only good photometry 
denoted by the flag {\it AAA} has been considered.}. The 3$\sigma$ limiting
sensitivity of this survey is 17.1, 16.4 and 15.3 mag for the three bands,
respectively. 

%______________________________________________________________
\section{The IRAC point source catalog}
\label{results}
\subsection{Positions of the IRAC sources}
\label{IRACpositions}
\begin{figure*}
  \centering
  \includegraphics[width=13cm]{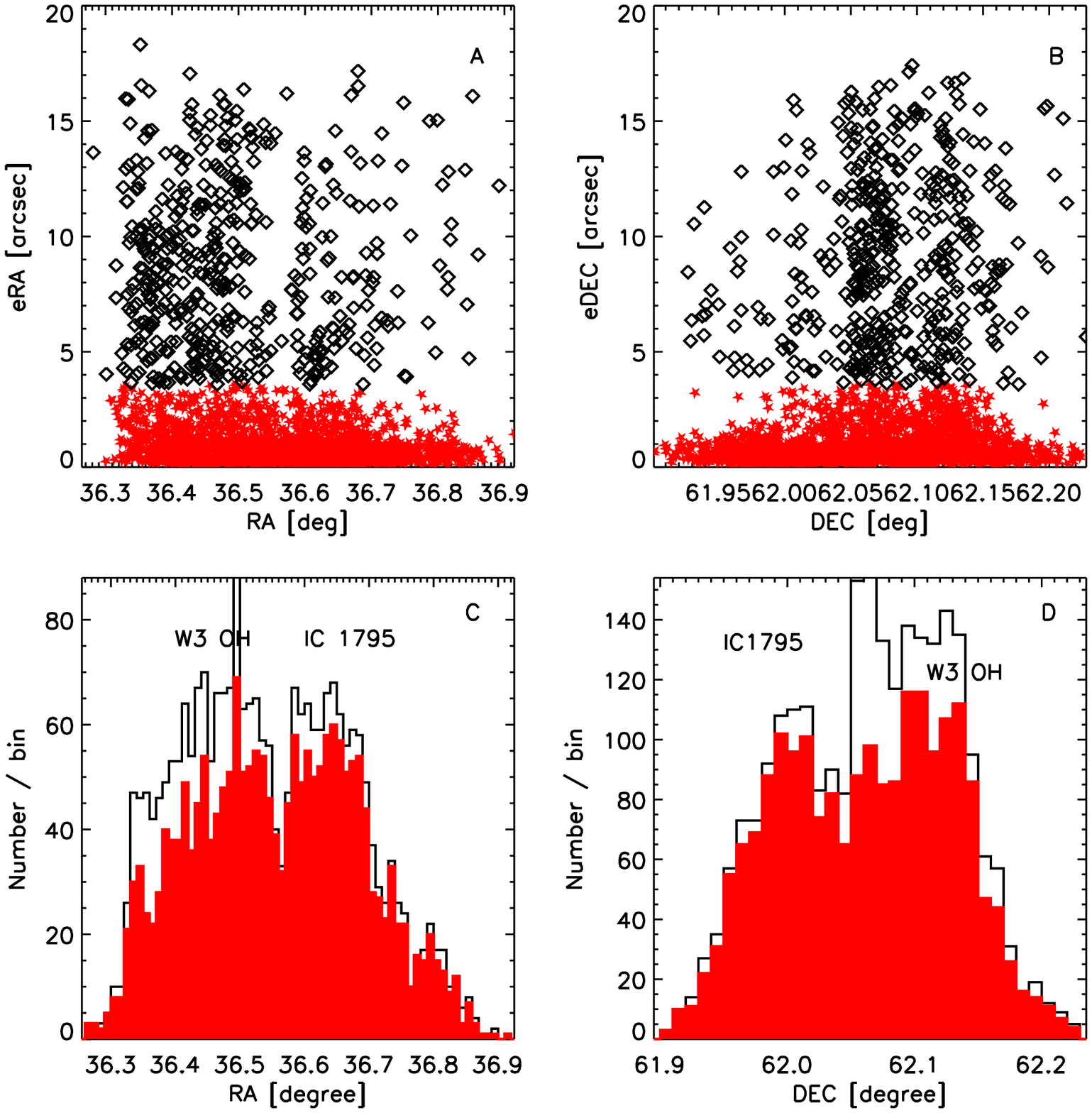}
  \includegraphics[width=6.5cm]{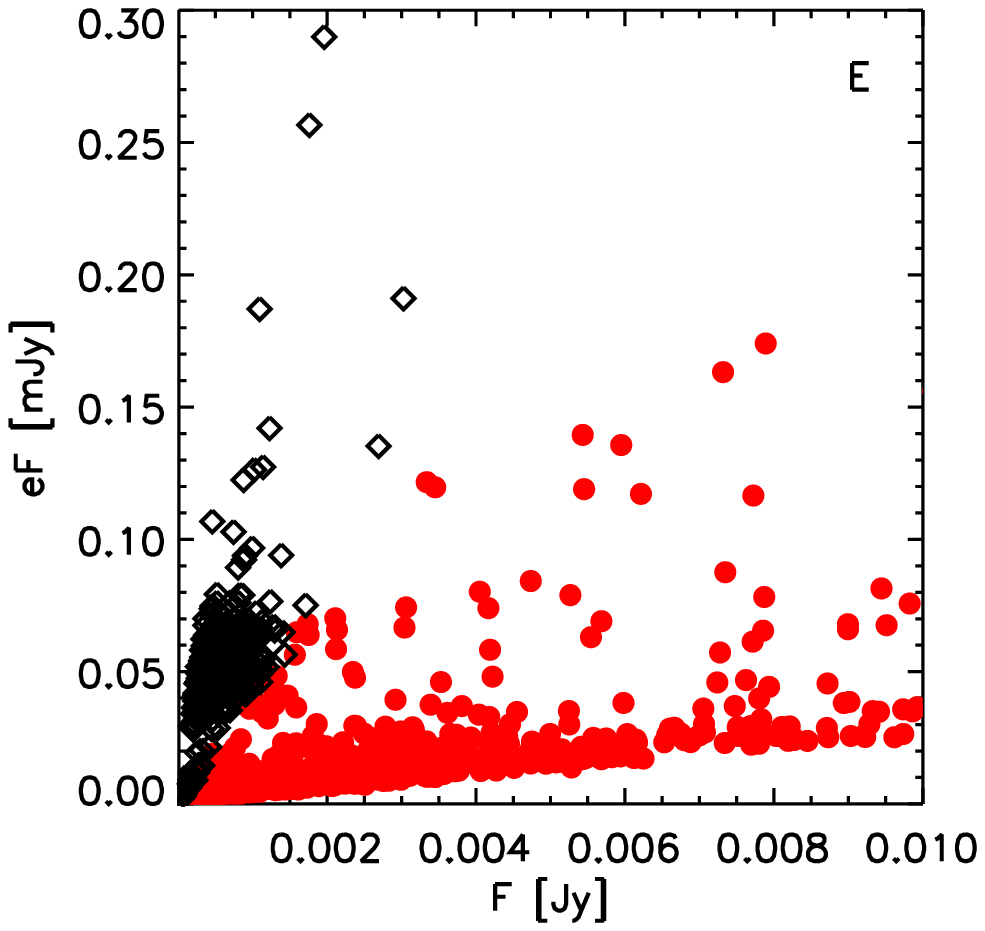}
  \caption{{\it A-B}: RA and DEC errors versus RA and DEC coordinates of all
    the sources detected in IRAC/ch1. The empty diamonds represent
    sources with coordinates errors larger than 3.6\arcsec. Sources with positional errors less 
    than 3.6\arcsec\, are highlighted with filled
    red stars. {\it C-D}: Black empty histograms represent the RA and
    DEC coordinates of all sources shown in panels A-B with empty diamonds. The
    red filled histograms represent RA and DEC coordinates of
    sources of our final lists (filled red stars in panels A-B). 
    {\it E}: Flux error versus the flux of the sources shown in
    the upper panels. The red filled stars represent the sources of our final
    lists selected with position errors in RA and DEC
    $\le$3.6\arcsec. Positional errors $>$3.6\arcsec (black empty diamonds)
     correspond to the fainter sources with larger positional errors. 
  }
  \label{coof}
\end{figure*}

\begin{figure*}
  \centering
  \includegraphics[width=13cm]{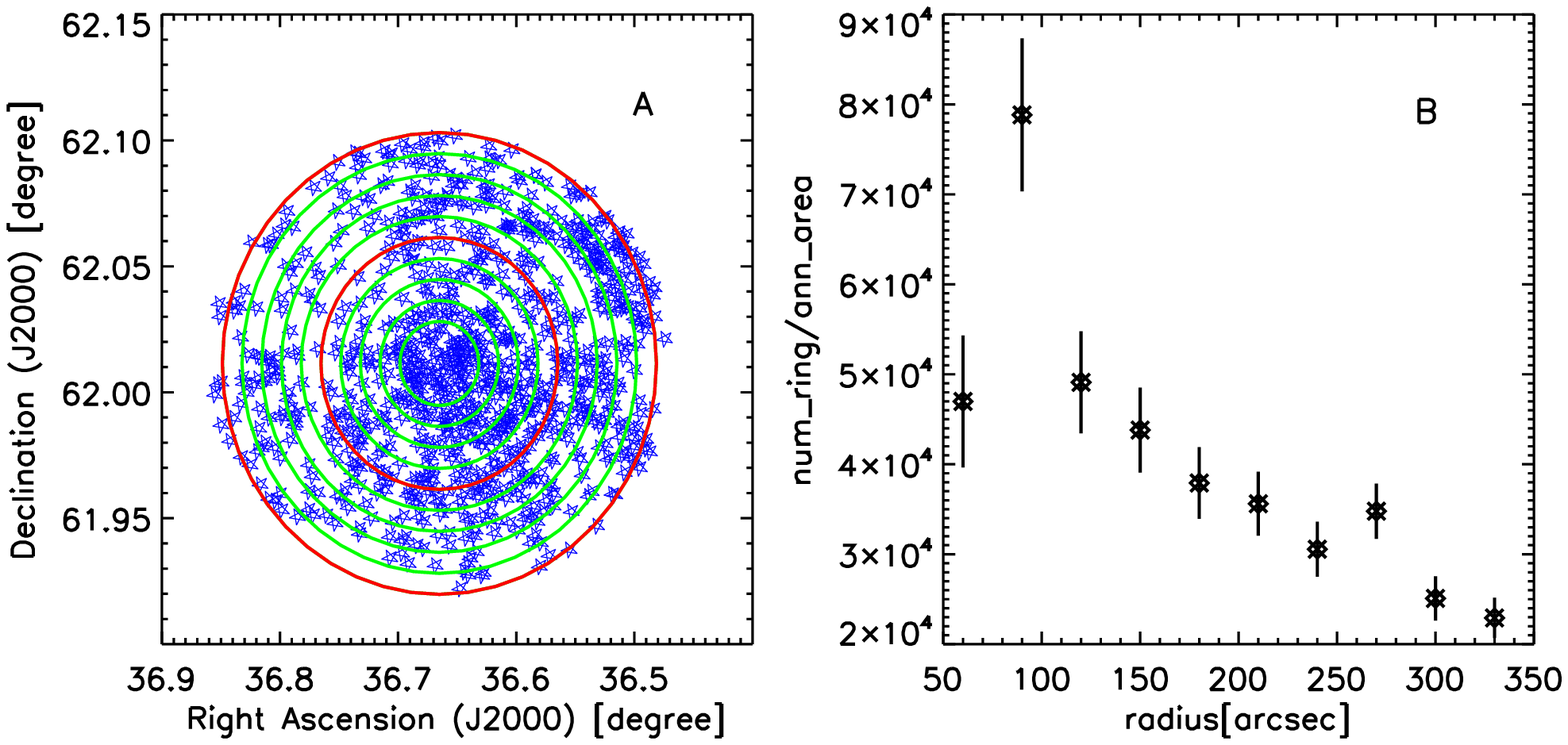}
  \includegraphics[width=13cm]{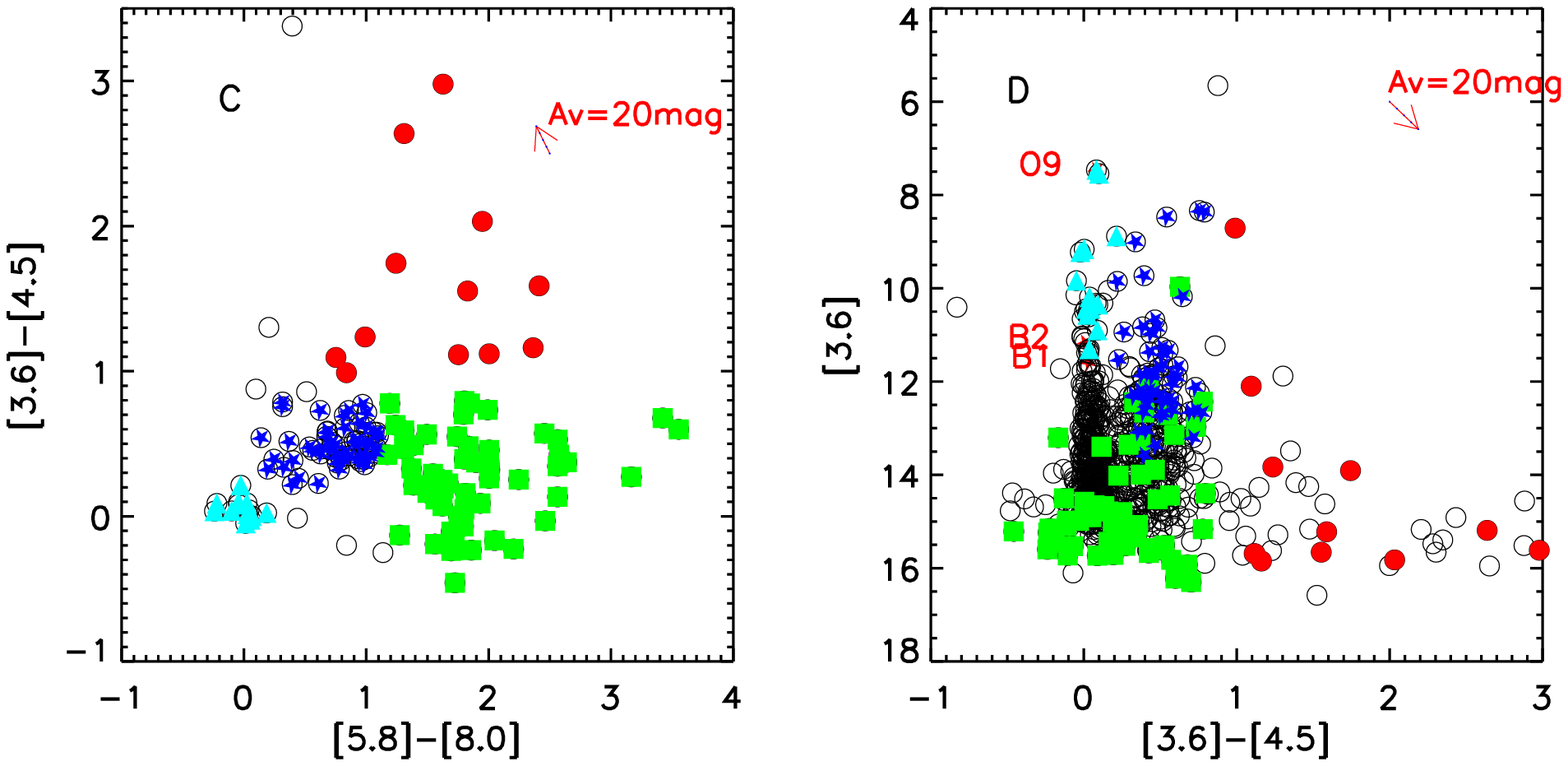}
   \caption{Spatial density distribution of all IR sources 
    ({\it A}), and their spatial density distribution as a
    function of the distance from the center of the cluster ({\it B}).
    The green circles represent radii from 60\arcsec\, from the center of
    the cluster to 330\arcsec\, with 30\arcsec spacing.
    {\it C}: Color-Color Diagram (CCD) of the IRAC/{\it Spitzer} sources
    detected within 330\arcsec\, from 
    the center of the cluster. The different symbols and colors represent the
    disk classification from \cite{Megeathetal2004} and \cite{Allenetal2004}: {\it filled (cyan)
      triangles:} photosphere/Class~III ; {\it filled (green) squares:}
    Class~I/II; {\it filled (blue) stars:} Class~II; {\it filled (red) circles:} Class~O/I; {\it D}:
    Color-Magnitude Diagram (CMD) of all the  
    sources identified at 3.6$\mu$m and 4.5$\mu$m. The different symbols
    represent the sources detected in four IRAC channels classified in the
    CCD on the right. {\it Empty (black) circles:} sources with magnitudes
    that do not follow the previous classification. The arrows represent an
    extinction A$_V$=~20~mag.} 
  \label{all}
\end{figure*}

The final photometric catalogs have been filtered for position errors, rejecting sources with positional errors in RA and DEC larger than 3.6\arcsec~(2 times the FWHM of the PSF in all IRAC channels). The distribution of source positions as a function of their error shows that sources with a larger position error are the fainter targets with a larger uncertainty in the flux (see Fig.~\ref{coof} for channel 1).\\
The histograms of the source positions in panels C-D of Fig.~\ref{coof} show a bimodal structure which reflects the distribution of stars in IC~1795 and W3-Main (ch1-ch3), and of stars in IC 1795 and W3-OH (ch2-ch4). From the peak of the histogram centered on IC 1795, the center of the cluster is found to be $(\alpha_{J2000}, \delta_{J2000})$ = (02h:26m:39s, +62$^{o}$:00$\arcmin$:41$\arcsec$).  This is $\sim 4$\arcmin\/ southeast of the cluster center chosen by \citet{Oeyetal2005}.\\
In panels A-B of Fig.~\ref{all}, the density distribution of all
infrared sources ({\it D$_{star}$}) is analyzed.  
{\it D$_{star}$} is computed as the total number of sources in a ring ({\it
  N}), defined between two consecutive radii from the cluster center, divided
by the ring area ({\it A$_{\rm ring}$}). 
%The circles are drawn each 30\arcsec\, in the range of
%60\arcsec\,-330\arcsec. 
 The annuli start at 60\arcsec\,from
the center of the cluster and end at 330\arcsec\,.
%\footnote{which corresponds to 3.2~pc, considering the cluster at $\sim$2~kpc }. 
The width of each annulus is 30\arcsec\,.
%In Fig.~\ref{distr_all}, these radii are in green, while in red the middle radius at 
%150\arcsec\, and the outer radius at 330\arcsec\, are highlighted.\\ 
The errors in the density distribution are computed as $\sqrt{N}/A_{\rm
  ring}$. 
The density distribution of sources 
detected at least in one IRAC channel has been analyzed as a function of the
distance from  the cluster center (panel A of Fig.~\ref{distr_all}).
$D_{star}$ peaks at 90\arcsec\, from the center of the cluster and
smoothly declines until 330\arcsec\, with the exception of a second minor peak at $\sim$270\arcsec\,
(panel B of Fig.~\ref{all}). \\
The density distribution is consistent with a cluster extension of $\sim$3.2~pc. The first peak in 
density distribution at $\sim$0.9~pc from the cluster center originates from a small clump of stars 
in the North West direction and by an asymmetric distribution of the stars in IC~1795.  

%% \begin{figure*}[h]
%%   \centering
%%   \includegraphics[width=10cm]{match_20may/flux_coo.ps}
%%   \includegraphics[width=10cm]{match_20may/coo.ps}
%%   \caption{{\it 1a-1b}: RA and DEC coordinates versus the RA
%%     and DEC errors of all the sources detected in IRAC/ch1.  {\it 1c-1d}:
%%     Flux percent error (eF/F) versus the flux of the of the sources shown in
%%     the upper panels. The red filled dots represent the sources of our final
%%     lists selected with position errors $\le$3.6\arcsec. Positions errors
%%     $>$3.6\arcsec (black filled dots) correspond to the fainter sources with
%%     larger errors,  (black filled in the lower panel).{\it 2a-2b}: Black empty
%%     histograms represent the RA and DEC errors of all filled black sources
%%     shown in panels 1. The filled histograms with red lines represent RA and
%%     DEC errors of sources of our final lists (filled red dots in panels 1).
%%     {\it 2c-2d}: As 2a-2b but for the RA and DEC coordinates. 
%%   }
%%   \label{cc1}
%% \end{figure*}
\subsection{Photometry of the IRAC sources}
\label{IRAC/Spitzer photometry}
The four {\it Spitzer} channels were matched using the IDL procedure {\it
  match\_xy.pro} in the IDL-based {\it Tools for ACIS Review and Analysis
  (TARA)} package\footnote{http://www.astro.psu.edu/xray/docs/TARA} \citep[][]{Broosetal2010}. 
  The source matching is based on agreement of positions
assuming that positional errors are 2-dimensional Gaussians at the 99\% confidence level.  \\  
The total number of point sources detected decreases as the IRAC wavelength
increases due to a decrease in sensitivity. 
In particular, within 330\arcsec\,  from the cluster center, we detect 918,
841, 303 and 243 sources at 3.6~$\mu$m, 4.5~$\mu$m, 5.8~$\mu$m and 8.0~$\mu$m
respectively. 143 objects were detected in all four IRAC channels.\\ 
The color-color diagram (CCD) of the sources detected in all four IRAC channels and
the color-magnitude diagram (CMD) of the sources detected at 3.6~$\mu$m and
4.5~$\mu$m (the most sensitive) are shown in panels C-D of Fig.~\ref{all}
(including the sources detected in all four bands). 
\cite{Megeathetal2004} and \cite{Allenetal2004} proposed a conversion scheme from IRAC colors to
source classification (Class~0,I,II,III sources), which has been later improved by \citet[][]{Gutermuthetal2008,
Gutermuthetal2009}. \\
The results of this
classification for our sources are summarized in Table~\ref{IR-X-table}.
 Using the CMD in panel C of Fig.~\ref{all}, we distinguish between
 sources with infrared excess (characteristic of circumstellar disks
around young star; [3.6]-[4.5]$\ge$0.2~mag) and ``photospheric sources''
 without infrared excess ([3.6]-[4.5]$<$0.2~mag). 
The {\it excess-threshold} adopted here ([3.6]-[4.5] = 0.2 mag) takes into 
account two factors: 1) model isochrones computed in the IRAC bands do not predict [3.6]-[4.5] colors exceeding 0.1~mag; see section 5.1 and Appendix~\ref{iso} for details; 2) the maximum uncertainty of our photometry 
(0.07 mag). 
Out of 592 sources detected at 3.6~$\mu$m and 4.5~$\mu$m, 327 show
an infrared excess and 265 are classified as ``photospheric'' sources.  
In the following sections we will call ``infrared sources'' objects detected at least in one {\it Spitzer} channel, and ``excess sources'' objects with  [3.6]-[4.5]$\ge$0.2~mag.\\
Out of 66 Class I and Class II sources previously classified using the definitions of \cite{Megeathetal2004} and \cite{Allenetal2004}, all sources have a [3.6]-[4.5]$\ge$0.2. Out of 57 ClassI\_II sources, 60\% have [3.6]-[4.5]$\ge$0.2. This ensures the reliability of the new scheme adopted to determine a Class I and Class II sources. It is important to notice that the effect of the visual exctintion is negligible at these wavelengths: following the extinction relations of \citet{Cardellietal1989} A$_{\rm V}$=2~mag (typical of a cluster at 2 kpc), corresponds to A$_{[3.6]}$=5.9$\cdot10^{-2}$mag. \\  
\cite{Ruchetal2007} analyzed the IRAC observations of the
entire W3 region available in the {\it Spitzer} archive (GTO PID 127) and obtained with a
total exposure time of 63.6~s in all IRAC channels. This corresponds to our
short exposure, and it is almost one order of magnitude shorter than our
long exposure time. 
They performed PSF photometry, using a version of DAOPHOT modified by
the Galactic Legacy Infrared Mid-Plane Survey Extraordinaire (GLIMPSE) team.
\\ 
 Within 330\arcsec\, from the center of the cluster, they identified 77
 sources visible in all four IRAC channels. With our deeper
 infrared survey we detected 143 sources. % (in all four IRAC
% channels simultaneously). 
There are however some differences between the two surveys: 
At 3.6$\mu$m and 4.5$\mu$m we do confirm 76/77 sources previously identified by
\cite{Ruchetal2007}, while 1 source has a  position error larger than
3.6\arcsec\, and it is not included in our analysis.  
At 5.8$\mu$m we confirmed the detection of 60 sources. Of the remaining 17
sources, 3 have position errors larger than 3.6\arcsec, while 14 have not been
detected by the PRF fitting method performed in APEX. 
At 8.0$\mu$m  we identified 32 sources. Of the remaining 45 sources,
26 have position errors larger than 3.6\arcsec\,, while 19 have not been
detected with APEX. 
We inspected all the non-matched detections (14 at 5.8$\mu$m and 19 at
8.0$\mu$m) in our deeper survey: In all cases we find that the local
background is highly variable compared to the source brightness and for this
reason we did not detect any point-source with our PRF fitting method. \\
 In summary we confirm 76 sources out of 77 sources detected by
\cite{Ruchetal2007} in the ch1 of IRAC and 32 sources in all four IRAC 
channels. In addition to these sources we identified 111 new sources in 
all four channels.  
\subsection{Extragalactic contamination of the IR catalog}
\label{IRcont}
The extragalactic contamination of the infrared source catalog is estimated 
using the 3.6~$\mu$m cumulative extragalactic counts from the IRAC/GOODS 
sample together with the incompleteness  correction from \citet{Franceschinietal2006}. 
About 10 galaxies  brighter than 1~mJy ($\sim$13.6~mag) at 3.6~$\mu$m are expected 
in an area of 1$^{\circ}$x1$^{\circ}$. The number of galaxies increases to about one order of 
magnitude up to $\sim$300 sources with fluxes higher than 0.1~mJy. Scaling these 
values for the IC~1795 area of about 0.26$^{\circ}$x0.26$^{\circ}$ the contamination of extragalactic 
sources is negligible (about 2.6 sources) down to 13.6~mag. In the 13.6 - 15.5~magnitude 
range, about 70 sources ($\sim$8\% of the total sources detected at 3.6~$\mu$m) might 
contaminate the IR source catalog of IC~1795. 
\section{Cluster membership}
\label{membership}
As outlined in the Introduction, a full understanding of protoplanetary disk evolution in rich stellar 
clusters requires a cluster membership sample identified in a disk-unbiased way. 
%To fully understand the evolution of protoplanetary disks in a large population
%of young stars, we need to study a sample that was previously identified in a
%disk-unbiased way.
Many photometric surveys of protoplanetary disks have been pursued with 
{\it Spitzer}, but {\it Spitzer} data alone suffers an important limitation: despite 
the excellent characterization of disks, sensitive IRAC surveys have biased 
the ratio of disked systems to non-disked systems in favor of the disk sources.\\
  Optical spectroscopy is widely used to identify cluster members in
  star-forming regions. This is mainly based on two spectral features:
the presence of H$\alpha$ emission at 6563\AA and strong absorption of the Li line at
  6708\AA. These diagnostics are powerful tools for identifying 
  young high accreting objects (H$\alpha$), or 
  %young stars but
  young low-mass stars with spectral type of K -- M (Li). 
  However, this method is not effective at identifying earlier spectral type members: G-type 
  stars, e.g., do not show a strong, age dependent Li I absorption and if they are non-accreting 
  they show only tiny or zero H$\alpha$ emission. An alternative,
  robust, approach to study the complete population of a young stellar cluster
  is to combine sensitive X-ray and infrared observations
  \citep[e.g.][]{Merceretal2009, stelzeretal2009, Wangetal2008,
    Getmanetal2009}. \\%PMS stars are indeed found to be highly active in the
%X-ray regime and are typically characterized by an infrared excess. \\
Sensitive X-ray surveys are effective in identifying pre-main sequence stellar
populations due to their enhanced magnetic flaring compared to older stars
\citep[e.g.][]{Feigelsonetal2007}. 
 Flux-limited X-ray samples have the advantage
of selecting young stars both with and without disks, but have the
disadvantage of missing stars with fainter bolometric luminosities and lower
masses.  X-ray selected samples are complementary to {\it Spitzer} samples because
they suffer less contamination from Galactic field stars or from diffuse
nebular emission than infrared surveys.   \citet{Getmanetal2009} provide details
on the relationships between stellar samples selected in {\it Chandra} and
{\it Spitzer} surveys of a star-forming region.
The high-spatial resolution telescope on the {\it Chandra X-ray Observatory}
is essential to resolve crowded regions such as the IC~1795 and the W3 complex. \\
For IC~1795 we classified as cluster members sources detected in both, our infrared and
  X-ray, surveys, recognizing that
extragalactic contamination may still be present (see discussion in section~\ref{xcontamination}). 
We also considered as potential members sources
  detected by {\it Spitzer} which show an infrared excess at 4.5$\mu$m but lack an X-ray
  counterpart.
 This approach is fundamental when studying distant young clusters ($>$1kpc), that can be strongly contaminated  by background and foreground stars. 
 
\subsection{Cluster membership based on X-ray detection}
\label{membershipX}
           {\it Chandra} source locations are matched to IRAC source locations using the
           IDL procedure {\it match\_xy.pro} in the TARA
           package\footnote{http://www.astro.psu.edu/xray/docs/TARA} \citep[][]{Broosetal2010}.
           Associations between {\it Chandra} and {\it Spitzer} sources are assumed to be real
           when the probability that the X-ray and infrared sources are coincident 
           exceeds 99\%, assuming bivariate Gaussian error distributions for the two
           source positions.  Within 330\arcsec\, from the cluster center, we find 280 
           associations between X-ray and infrared sources, and 9 cases where a 
           single IRAC source has two or more possible {\it Chandra} counterparts. 
           %We did not find a single {\it Chandra} source with two or more
           %possible IRAC counterparts. 
           The {\it Chandra-Spitzer} matched positions are
           listed in Table~\ref{irac-x-pos}. The corresponding optical, 2MASS, and
           IRAC photometry is compiled in Table~\ref{irac-x}. \\
           We analyzed the distribution of infrared sources with X-ray counterpart as a
           function of the distance from the cluster center (panels A-B of Fig.~\ref{distr_all}).
           The distribution peaks at $\sim$90\arcsec\, from the
           center, declines down to 200\arcsec\, and then remains
           constant to 330\arcsec\, (except for a second peak at
           $\sim$270\arcsec). 
           The positions of these two peaks are also found in the distribution of infrared 
           source positions (section~\ref{IRACpositions}, panel B of 
           Fig.~\ref{all}).\\
           %This is similar to what we have noted previously
           %(section~\ref{IRACpositions}, panel B of Fig.~\ref{all}).\\
           The number of X-ray sources alone, with and without infrared counterpart, and
           infrared sources (e.g. detected at least in one {\it Spitzer} channel) without X-ray 
           counterpart are summarized in Table~\ref{IR-X-table}. 
           The final catalog of the IC~1795 cluster members is in Table~\ref{irac-x}. 

\subsection{Contaminants in the X-ray catalog}
\label{xcontamination}
%In this section we discuss the sources of contamination of an X-ray survey. 
The sensitivity limit to X-ray sources in the IC~1795 cluster using
the {\it ACIS Extract} methods is approximately $L_x \sim 5 \times
10^{29}$ erg s$^{-1}$ in the {\it Chandra} total band, $0.5-8$ keV,
assuming moderate absorption around $N_H \sim 10^{21}-10^{22}$ cm$^{-2}$
($A_V\sim0.5-5$mag) and source spectra typical of PMS
stars.  This is not a well-established value due to differences in
source spectra, spatial variations in extinction, loss of
sensitivity off-axis due to degradation of the {\it Chandra} point
spread function, and gain of sensitivity in some regions due to
overlapping exposures in the W3 ACIS mosaic.  
In this paper we use only the X-ray positions. A more detailed analysis of
the X-ray observations will be presented in a forthcoming paper (Townsley et
al.~in preparation). We estimate the completeness limit to be around $L_x \sim
1 \times 10^{30}$ erg s$^{-1}$.  Based on the well-established empirical
correlation between X-ray luminosity and stellar mass \citep{Telleschietal2007},
this X-ray limit corresponds to a mass completeness limit of about 1~M$_\odot$, although a considerable number of lower mass stars will be included in the $Chandra$ sample.\\
%this X-ray limit corresponds to a mass limit of around 1 M$_\odot$. \\
% Thus, for a standard Initial Mass Function (Chabrier 2003), the
% X-ray sample should detect about 5\% of the entire stellar pre-main
% sequence population (including brown dwarfs).   
A fraction of the X-ray sources should be contaminants, uniformly distributed,
unrelated to the IC~1795 cluster \citep[see discussion in][]{Getmanetal2006}.
About $\sim30-40$ faint and heavily-absorbed X-ray sources in a 100
arcmin$^2$ region (which is of the order of the cluster size) will be
background extragalactic objects, mainly quasars and AGNs, 
seen through the molecular material in the W3 complex and Galaxy along the
line-of-sight. 
At least 20-30 lightly absorbed X-ray sources in a 100~arcmin$^2$ region should be foreground Galactic field stars, and $\sim 10$ sources should be background stars. The value depends on the amount of recent star formation along the line-of-sight to the W3 complex.\\
%Roughly $20-30$ lightly-absorbed X-ray sources in a
%100 arcmin$^2$ region should be foreground stars and $\sim10$ sources background stars.\\
We expect that most of these contaminants, except for foreground stars, will not have an
infrared counterpart and so will not contribute to our source list. 
These extragalactic and Galactic contaminants in the X-ray sample will exhibit a
random spatial distribution, as is clearly shown for X-ray sources without
infrared counterpart in panels E-F of Fig.~\ref{distr_all}.  
However a few quasars can have IRAC colors and X-ray fluxes similar to CTTS \citep[e.g.][]{Richardsetal2009}. 
This last source of contamination can only be removed via optical spectroscopy.

\subsection{Cluster member IR properties}
\begin{figure*}
 \centering
  \includegraphics[width=13cm]{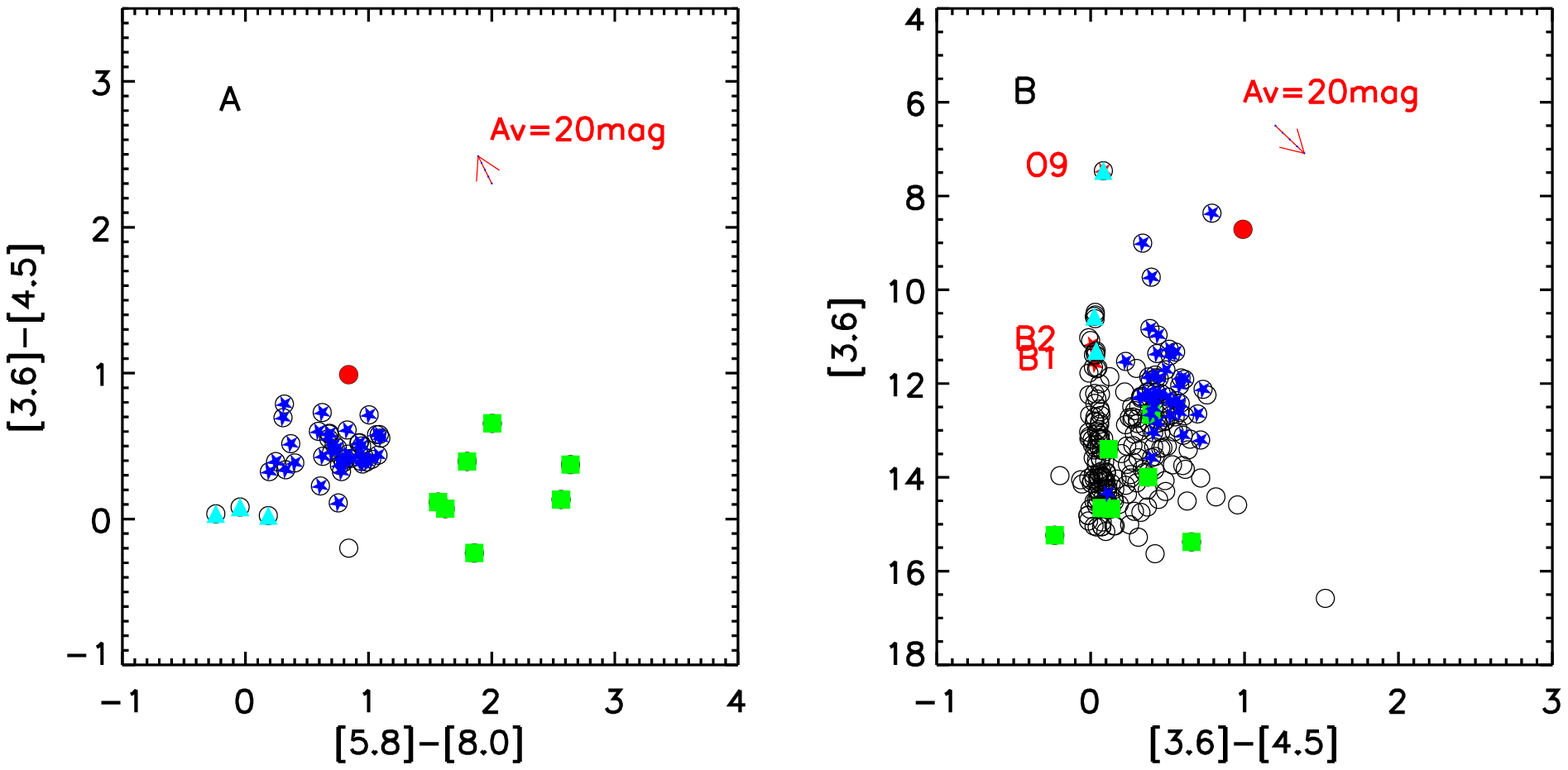}
  \includegraphics[width=13cm]{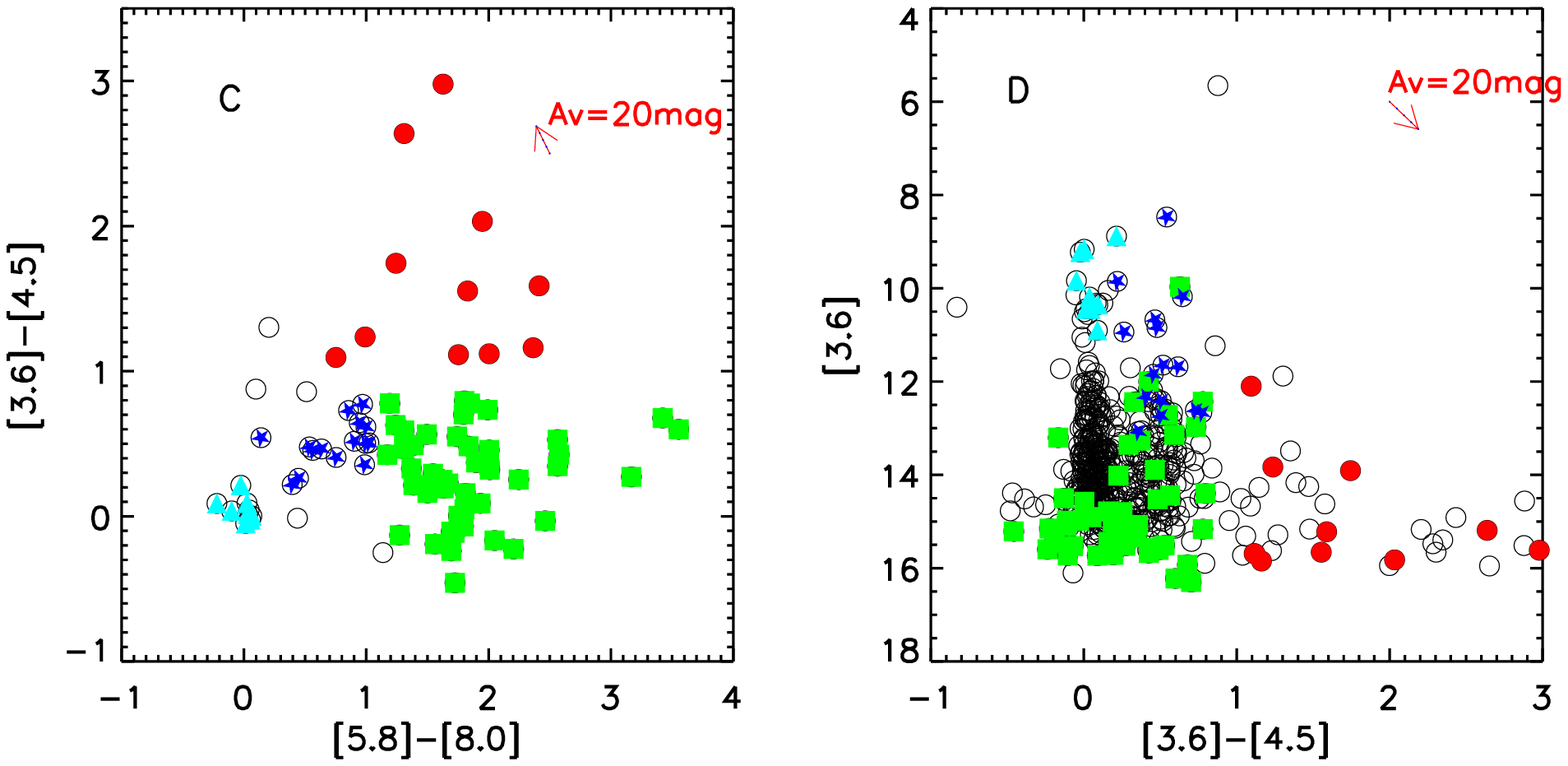}
    \caption{{\it Left}: CCD of the IRAC/Spitzer sources
      detected within 330\arcsec\, from the center of the cluster (empty black
      circles). The different symbols and colors represent the disk
      classification from \cite{Megeathetal2004} and \cite{Allenetal2004}: {\it filled 
      (cyan) triangles:} photosphere/Class~III ; {\it filled (blue) stars:} Class~II; {\it filled (green) squares:}
      Class~I/II; {\it filled (red) circles:} Class~0/I; 
      {\it Right}: CMD of all sources identified in
      IRAC/ch1 and IRAC/ch2. The different symbols represent the
      sources detected in the 4 IRAC channels classified in the CCD
      on the right. The arrows represent an extinction A$_V$=~20~mag. {\it A-B:}
      infrared sources with an X-ray counterpart. {\it C-D:} infrared sources
      without an X-ray counterpart.
            }
      \label{cc+am_all}
 \end{figure*}

\begin{figure*}
  \centering
  \includegraphics[width=13cm]{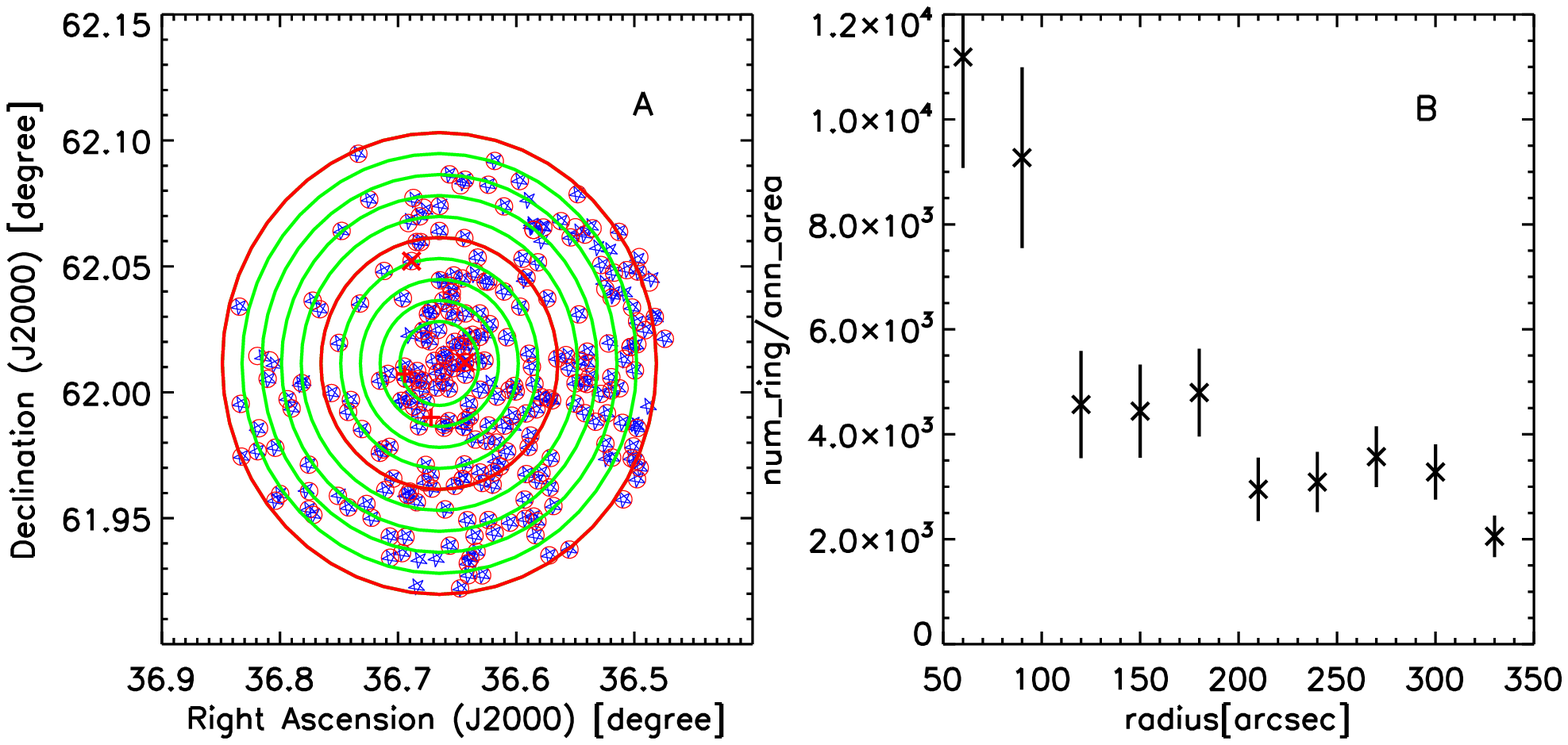}
  \includegraphics[width=13cm]{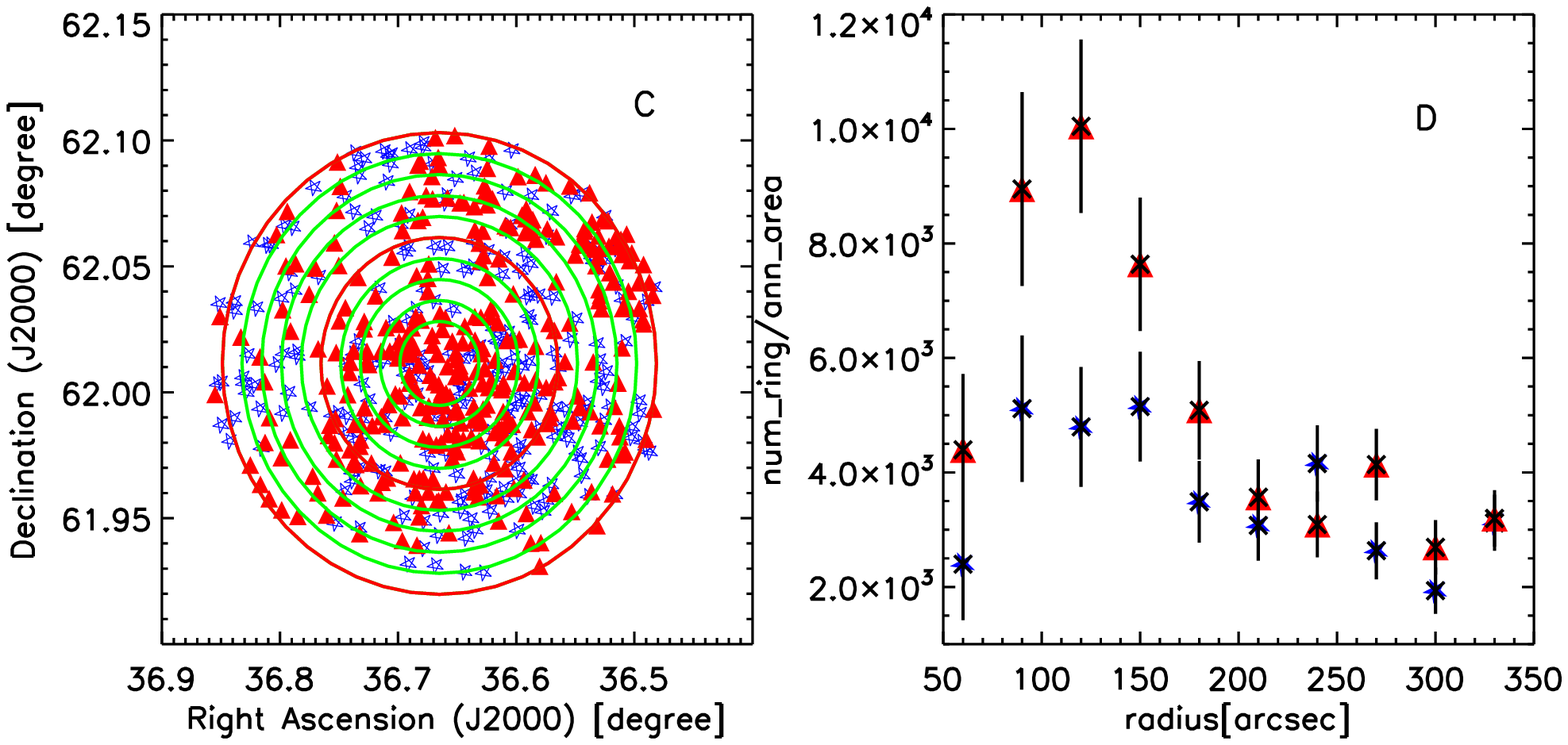}
  \includegraphics[width=13cm]{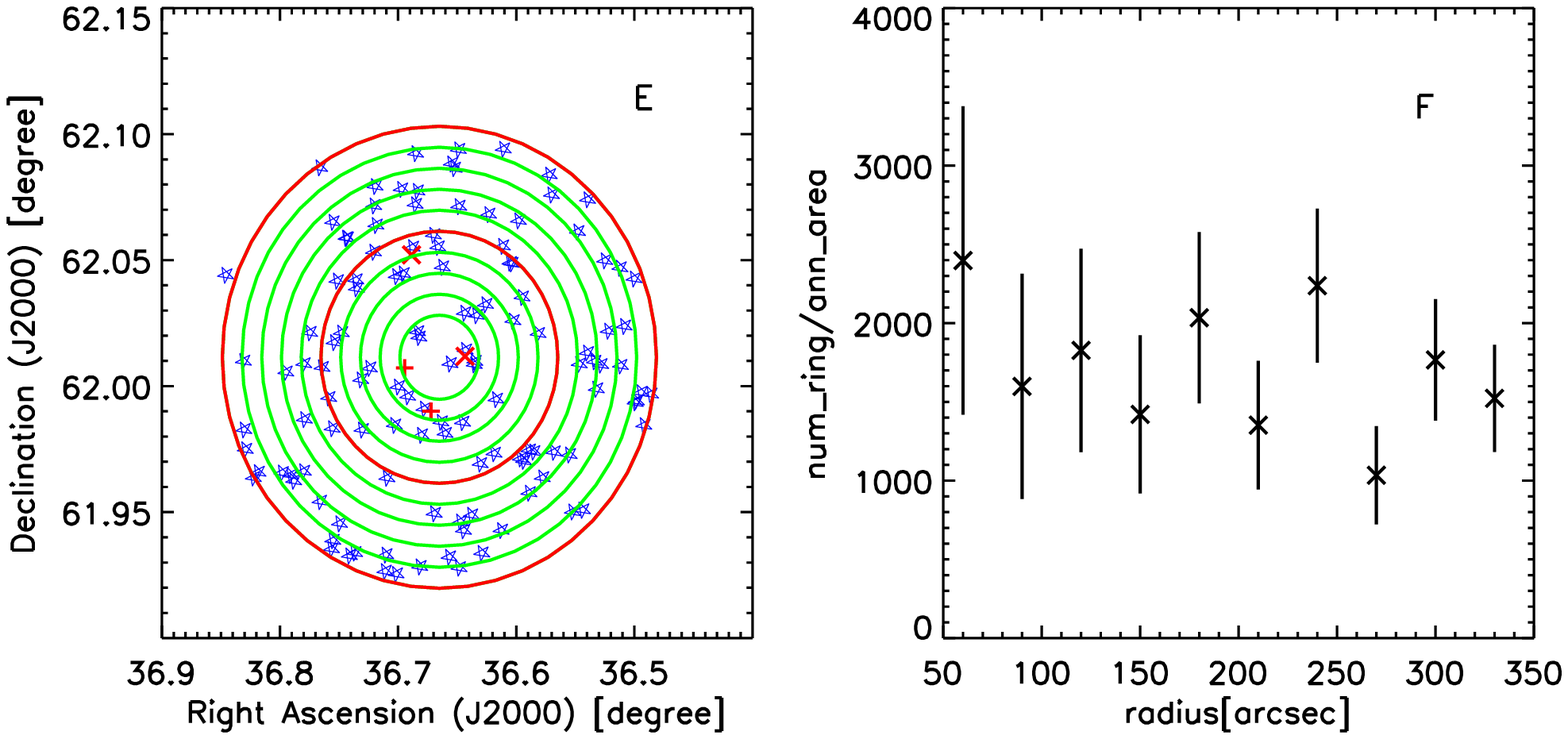}
  \caption{Spatial density distribution of all the IR sources 
    ({\it left column - A, C, E}), and their spatial density distribution as a
    function of the distance from the center of the cluster ({\it right column
    - B, D, F}).
    The green circles represent annuli from 60\arcsec\, from the center of
    the cluster to 330\arcsec\, with a spacing of 30\arcsec. 
    The 'x' and '+' symbols represent the positions of the O and B stars in the cluster. 
    %{\it 1:} sources detected
    %in all the IRAC channels within 330\arcsec from the center of the
    %cluster. 
    {\it A-B:} infrared sources with an X-ray counterpart. {\it C-D:} infrared
    sources without an X-ray counterpart: the triangles represent sources with
    excess and the stars sources without excess. {\it E-F:} X-ray sources
    without infrared counterpart.} 
  \label{distr_all}
\end{figure*}

%candidates and IR contamination}
\label{candidates}
In section~\ref{membershipX} the cluster members have been defined as an infrared source with 
a X-ray counterpart. 
In panels A-B of Fig.~\ref{cc+am_all} the CCD and CMD of
the cluster members (i.e. infrared sources with X-ray counterpart) detected within 330\arcsec\, from the center of the
cluster are shown. We distinguish between cluster members with infrared excess
([3.6]-[4.5]$\ge$0.2) and ``photospheric'' sources without infrared excess
([3.6]-[4.5]$<$0.2). \\
%Out of 837 sources detected in ch1 and ch2, we find 479 IRAC excess sources\footnote{
%defined as [3.6]-[4.5]~$\ge$~0.2}. \\
In panel A of Fig.~\ref{cc+am_all} the 145 sources detected in all IRAC
channels are shown. In Table~\ref{IR-X-table} we summarized the classification of the disks
  using the IRAC colors from \cite{Megeathetal2004} and \cite{Allenetal2004}. \\
  For comparison in panels C-D of Fig.~\ref{cc+am_all} we plot the IRAC detections with no X-ray
  counterpart: we notice that sources with and without infrared excess in the CMD of panel D reach $[3.6] \sim 16$~mag, while in the CMD of panel B sources without excess have 3.6$\mu$m magnitudes down to 15.5~mag, and only few sources with excess have 3.6$\mu$m magnitudes larger than 14~mag. 
The density distribution\footnote{defined as in section~\ref{IRACpositions}} of the cluster members is shown in panels A-B of  Fig.~\ref{distr_all}. The distribution peaks within 100$\arcsec$ from the center of the cluster. 
In panels C-D of Fig.~\ref{distr_all} the density 
distribution of infrared sources without a X-ray counterpart is shown. 
Out of 593 sources detected in ch1 and ch2 without a X-ray counterpart, we find 248 IRAC 
excess sources\footnote{defined as [3.6]-[4.5]~$\ge$~0.2}. 
The distribution of sources with excess is similar to the distribution of the
cluster members, while the flat distribution of sources without excess suggests that 
 most of them are not related to the cluster. 
 For this reason, infrared sources without X-ray counterpart but with excess emission have been classified as 
{\it cluster member candidates}\footnote{see section~\ref{IRcont} for a possible 
extragalactic contamination }. 
Table~\ref{onlyirac-withexc-pos}-\ref{onlyirac-withexc} give positions and IRAC fluxes of the cluster member candidates (i.e. infrared sources with excess without X-ray counterpart). %3$\sigma$ upper limits of the 340
%non-X-ray infrared excess sources. The standard deviation ($\sigma$) of the
%background was computed within a 10x10 pixel box centered on the X-ray position.\\

\section{Age and Mass Distribution of the cluster members}
\label{analysis}
In this section we estimate the age, the mass distribution and the completeness
 of the cluster members of IC~1795, defined in section~\ref{membership} as infrared sources with an X-ray counterpart.
\subsection{Age of the cluster}
\label{age}
\begin{figure}
  \centering
   \includegraphics[width=14cm]{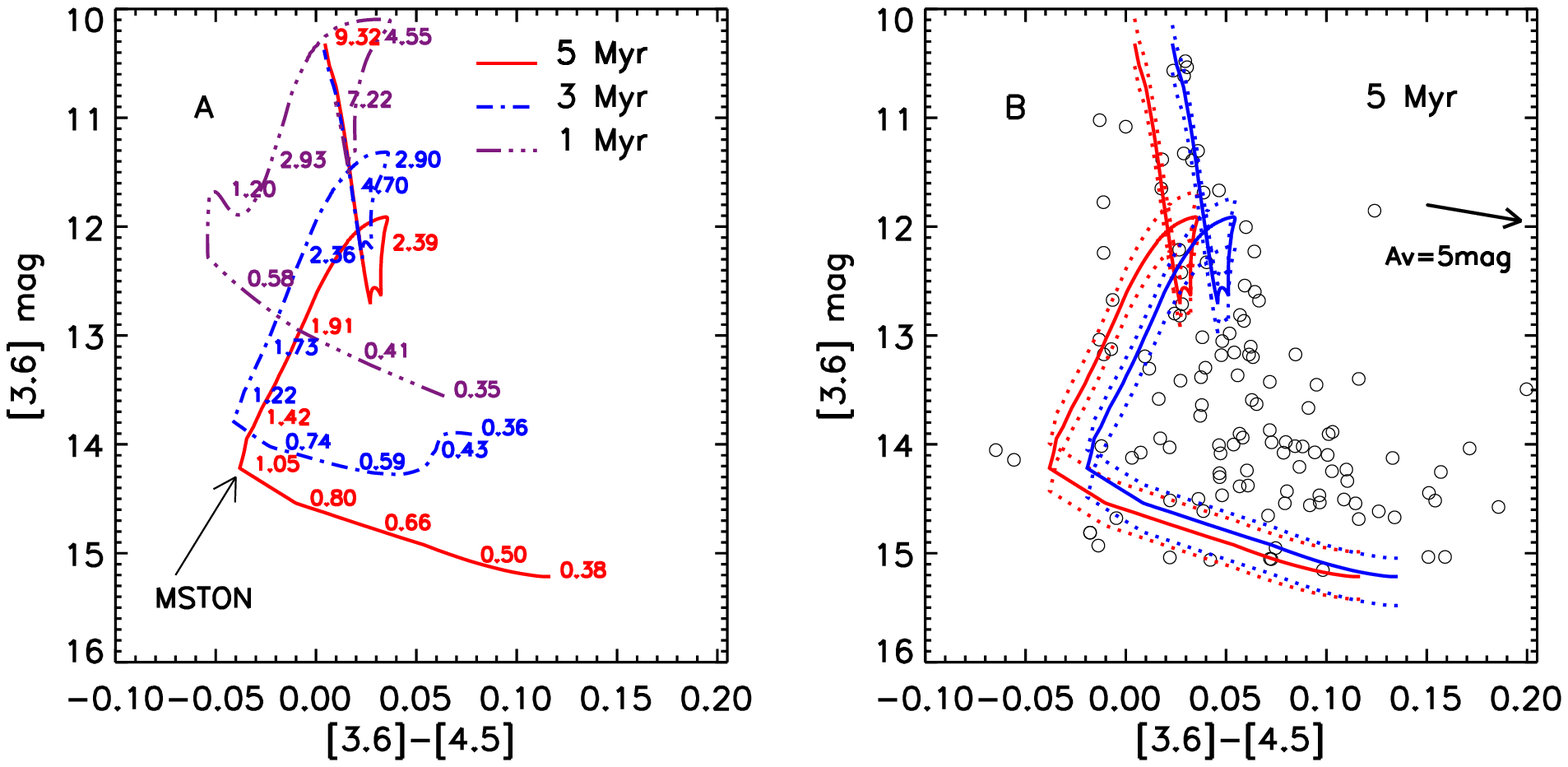}
       \caption{ {\it A:} 1~Myr, 3~Myr and 5~Myr isochrones computed in the IRAC magnitudes, assuming a
     cluster distance of 2~kpc  and Av = 2 mag. Overplotted are the position of the different masses over each isochrone. The arrow shows the position of the MSTON (Main Sequence Turn-On) for the 5~Myr isochrone. {\it B:} The circles represent the data (as in Fig.~\ref{age}B) and overplotted is the 5~Myr isochrone reddened by Av = 2 mag and Av =  4 mag. The dotted lines represent the 5~Myr isochrones but shifted at a maximum
    and minimum distance of 2.2 and 1.8~kpc (as in Fig.~\ref{age}B). 
       }
       \label{isoc}
 \end{figure}

 The ages of PMS associations are usually estimated by comparing the location of 
 the association members in a CMD or Hertzsprung-Russell diagram, to isochrones 
 resulting from the predictions of PMS evolutionary model grids. This comparison 
 usually makes use of dereddened optical or near-infrared photometry unaffected by the presence 
 of a protoplanetary disk, but optical spectroscopy of most of the cluster 
 members, to quantify the differential reddening to each source, is not available. The positions of the sources in these CMDs can thus be either affected by differential extinction, or by a spread in age, as well as by undetected binarity.\\
 However, for IC 1795 members we do have extensive 2MASS J-band and 
 IRAC infrared [3.6] and [4.5] photometry. Thus to optimize the isochronal fitting 
 of IC 1795 members, we first look for the best magnitude and color to distinguish between sources with and without infrared excess. This is needed to clean the CMD of the infrared-excess sources.\\  
The {\it J} magnitude is not affected by infrared excess (which in the 
different classes of disks would usually start, at least, from the {\it H}-band), and is 
sensitive to the stellar mass. The first CMD analyzed used the {\it J} vs. {\it J} - {\it H} 
magnitudes (Fig.~\ref{age}A). However the
disadvantage is that the {\it J} and {\it H} magnitudes are strongly affected by extinction
along the line of sight. This represents a serious problem in particular working on a 
cluster at more than 1~kpc in distance such as IC~1795. Overlaid on
the CMD in Fig.~\ref{age}A are the 3~Myr and the 5~Myr isochrones obtained using the FRANEC
evolutionary code (see Appendix~\ref{iso}). 
The isochrones are reddened with $A_V$ = 2~mag which is the minimum extinction
expected for a cluster at distance of 2~kpc. Without any information on a possible
differential extinction within the cluster, from this
CMD it is not possible to distinguish between the different isochrones. The positions
of the sources can be either affected by additional extinction, or by a
spread in age, as well as by undetected binarity.\\ 
Different CMDs, combining in different ways the {\it J}, 3.6$\mu$m
and 4.5$\mu$m magnitudes, have been further compiled.   
The advantage of using the J-[3.6] or J-[4.5] combinations is that photospheric
and infrared excess sources are clearly separated in color. % (see the module of the extinction $A_V$ in Fig.~\ref{irac-j}). 
The advantage of using the [3.6]-[4.5] combination instead, is that at these 
wavelengths the effect of interstellar extinction is lower (see the module of the
extinction $A_V$ in Fig.~\ref{age}). 
We thus use the [3.6] vs. [3.6]-[4.5] CMD to distinguish between objects with and
without infrared excess, taking advantage of the negligible extinction at these wavelengths. \\
%Figures~\ref{cm-jhk}-\ref{opt} show the CCDs and CMDs in the near-infrared and
% optical planes, respectively. %The red and the
%%blue empty circles are sources classified with and without infrared excess in
%%the mid-infrared CMD of panel 4 of Fig.~\ref{irac-j}. 
%The filled circles with different shapes represent the different classes of
%disks (see panel A of Fig.~\ref{cc+am_all} for the symbol key). Overlaid on
%the diagrams are the 3~Myr and the 5~Myr isochrones obtained using the FRANEC
%evolutionary code (see Appendix~\ref{iso}). %\citep{Chieffietal1989,Deglinnocentietal2008}. 
%We reddened the isochrones with $A_V$ = 2~mag which is the minimum extinction
%expected for a cluster at distance of 2~kpc. Without any information of possible
%differential extinction within the cluster, from these
%CMDs it is not possible to distinguish the different isochrones. The positions
%of the sources can be either affected by additional extinction, or by a
%spread in age, as well as by undetected binarity.\\  
%The best choice is to use the isochrones computed in the IRAC colors and compare
%them to the sources without infrared excess as shown in Fig.~\ref{age}.
%Given the independent estimate of the distance of 2.0 kpc \citep[e.g.][]{Xuetal2006}
%and given that in the IRAC bands the foreground extinction for our cluster is
%negligible, the only parameter that determines the vertical position of the
%Main-Sequence Turn-On ({\it MSTON}; see appendix~\ref{iso}) is the age of the cluster.
In Fig.~\ref{age} panel B we show the [3.6] - [3.6]-[4.5] CMD of the Class~III cluster members, together with the 1~Myr, 3~Myr and 5~Myr isochrones. 
In this CMD, the only parameters that determine the vertical position of the
Main-Sequence Turn-On ({\it MSTON}; see appendix~\ref{iso}) are the age and the distance of the cluster.  We notice that at these wavelengths an error of 10\% in distance corresponds to an uncertainty of less than 1~Myr in age and the reddening is unimportant because the reddening 
vector (see Fig.~\ref{age} panel B) runs parallel to the low-mass population. We also investigate whether an additional extinction of 2 mag in V would change the position of  {\it MSTON}, and hence the conclusions about the upper limit on the age of the system. In Fig.~\ref{isoc}B the 5~Myr isochrone is reddened by Av = 2 mag and Av = 4 mag. Taking into account the errors on the 
distance of the cluster (shown also in Fig.~\ref{isoc}B) and on the photometry of the single sources (the maximum error was 0.07 mag; see section~\ref{IRAC/Spitzer photometry}), we 
conclude that, within the errors, all the stars still lie close or above the 5~Myr isochrone.  
For these reasons the 
apparent age of the cluster can be determined using reddened photometry. 
From the {\it MSTON} we can conclude that the cluster is younger than 5~Myr and 
it  likely has an age between 3 and 5~Myr. This is consistent with an earlier
estimate \citep{Oeyetal2005} which was obtained from the position in the H-R diagram 
of the early type sources of IC 1795. \\
Due to the presence of other stars with a 3.6$\mu$m magnitude in the range of 
12.5~-~14~mag which do not follow the shape of the isochrones, we investigate the
possibility of a spread in age of the cluster members. 
We first notice that the 1~Myr isochrone does not reproduce the distribution of
the observed sources on the CMD (Fig.~\ref{age}).  
If we presume that the cluster has a spread in age between 1 and 5 Myr due to different episodes of star formation, sources with a 3.6$\mu$m magnitude in the range of 12.5 - 14 mag would represent the younger population of the cluster, since for the 1~Myr isochrone all stellar masses reside within this magnitude range (Fig.~\ref{age}). 
In this case, they would also have a different spatial distribution in respect to the older population
with 3.6$\mu$m$>$14mag. 
However, the random spatial distribution of the photospheric sources with different [3.6] mag argues against a significant spread in age in the cluster population caused by triggered episodes of star formation. 
A second argument against a young population present in the cluster regards the CMD using the 2MASS magnitudes (Fig.~\ref{age} panel A). The overlaid isochrones are shifted 
 assuming a distance to the cluster of 2 kpc and Av = 2 mag, which is the minimum extinction for 
a cluster at that distance. 
In this CMD the bulk of the sources are bluer and fainter than the 1 Myr isochrone.\\
These arguments, however, do not exclude that a younger population might be present in the cluster and is viewed in projection over the older one. \\
A possibility is that stars with $[3.6] = 12-14$ % 3.6$\mu$m = 12-14~mag
are binaries.  
In section~\ref{secmassfunc} we test this hypothesis by including the binarity in
a simulation of the theoretical luminosity function of the cluster.  
%% \begin{figure*}
%%  \centering
%%   \includegraphics[width=15cm]{match_20may/IC1795_nir_circ11_1.ps}
%% %  \includegraphics[width=8cm]{match_20may/IC1795_nir_circ11_2.ps}
%%    \caption{Near infrared CCD ({\it right panel}) and
%%      CMD ({\it left panel}) of the cluster member sources
%%      with a near-infrared 2MASS counterpart. The {\it `crosses'} on the
%%      reddening vectors show A$_V$= 2, 4, 6, 8~mag. The 5Myr and ({\it dashed
%%        red line}) and the 3Myr ({\it solid blue line}) are reddened
%%      using $A_V=~2$.
%%             }
%%       \label{cm-jhk}
%%  \end{figure*}
%% 

%% \begin{figure}
\subsection{Luminosity and Mass Functions of IC~1795}
\label{secmassfunc}
\begin{figure*}
  \centering
   \includegraphics[width=13.2cm]{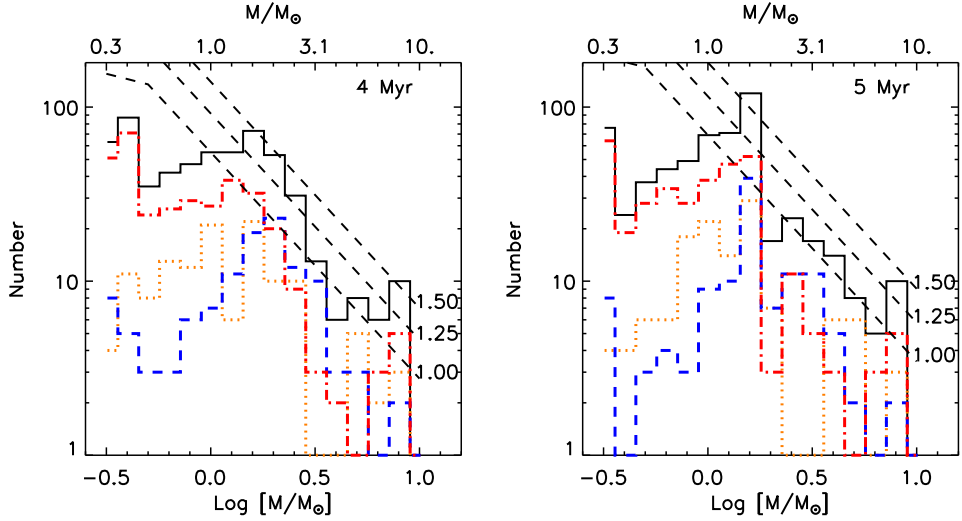}
\caption{
     Mass function computed at [3.6]~$\mu$m of the cluster members (infrared sources with X-ray counterpart) and candidates (infrared sources with excess without X-ray counterpart) using isochrones at 4 and
    5~Myr. Overplotted are the galactic IMFs \citep{Kroupa2002} normalized at
    different completness masses: 1~M$_\odot$,1.25~M$_\odot$,1.5~M$_\odot$. %,2~M$_\odot$.   
   The mass function at [3.6]~$\mu$m of the different subgroups are overplotted in different line-styles and colors: the {\it red dashed-dotted} line represents the candidates members alone (i.e. Spitzer sources with excess); the {\it blue dashed} line represents the cluster members with excess (i.e. Spitzer sources with excess with X-ray counterpart);  the {\it orange dotted} line represents the cluster members without excess. 
   }
       \label{massfun}
  \end{figure*}

The comparison of the observed and theoretical luminosity function enables to estimate the completeness limit of our {\it Spitzer-Chandra} survey.\\
%Fig.~\ref{completness} (panel~1) shows the observed luminosity function
%at [3.6] of the cluster members without infrared excess.
%Panels 2 and 3 of Fig.~\ref{completness} show respectively the 3.6$\mu$m
%magnitude distribution of the cluster members with a near-infrared and
%optical
%counterpart. 
%The different distributions suggest that the optical and near-infrared
%surveys are not as deep as our IRAC survey ().
The observed luminosity function is compiled using the 3.6$\mu$m magnitude as
a tracer of the stellar photosphere. 
%The fraction of sources with infrared excess may depend on the stellar mass 
%(as investigated in section ~\ref{diskfrac-mass}). For this reason the
%distribution of the cluster members without infrared excess is not
%representative of the real luminosity function. 
To have a description of the entire stellar population of IC~1795, the observed luminosity function includes cluster members with and without infrared excess and cluster member candidates (defined in section~\ref{candidates}). 
 This approach is necessary because the disk fraction may be mass dependent (section~\ref{discussion}).
We used the method described in Appendix~\ref{diskfrmidir} to determine the ``photospheric'' 3.6$\mu$m magnitude of the sources with infrared excess. The same has been done for the cluster member candidates.\\
 We computed the theoretical luminosity function using a galactic initial mass
 function (IMF) \citep{Kroupa2002} and the 4~Myr and 5~Myr isochrones respectively. 
The simulated population is normalized to the observed stellar sample down to a certain completeness magnitude limit. 
% The
 %number of simulated stars is normalized in such a way that in each simulation
 %we have the same total number of simulated and observed stars below a certain magnitude which represents the completeness of our survey.
%the limiting magnitude of [3.6] = {\bf 15.5~mag}. 
%These numbers are computed on the
 %observed luminosity function which includes the cluster members
 %candidates. This is the limit where we expect to have 100\% completeness in our sample.\\ 
We compare the simulations obtained using a completeness value of 13.5~mag, 13.75~mag, 14~mag, 14.25~mag, 14.5~mag and 15~mag at 3.6$\mu$m. 
We conclude that the completeness of our survey at 3.6$\mu$m is 14~mag. \\
We notice, however, that for both ages of 4 Myr and 5 Myr, the simulations do not well-reproduce the number of sources in the 3.6$\mu$m magnitude range of $13-14$. This might be an effect of a binary population unresolved in the cluster. To test this, we ran simulations with 30\% and 50\% binary fractions, with primary and secondary masses drawn independently from the same IMF. 
The outcomes are only slightly different from our original calculations. This implies that the luminosity function is only marginally affected by binarity.\\
In order to convert the completeness expressed in magnitude into the corresponding stellar mass value, the mass function of IC~1795 is now extracted.\\
Using the 4~Myr and 5~Myr isochrones, from the observed luminosity function we derived the stellar masses and hence construct the mass function for the cluster.  
We used the method described in Appendix~\ref{mag-mass} to assign a stellar mass to each source. 
The resulting observed total mass function in shown in Fig.~\ref{massfun} together with the mass functions of the single sub-groups considered for the total mass function: the cluster members\footnote{Spitzer sources with X-ray counterpart} without and with infrared excess and the cluster candidates\footnote{Spitzer sources with excess without X-ray counterpart}. The theoretical IMF  \citep{Kroupa2002} 
is normalized to the total number of stars down to 14~mag.  We notice that all the distributions peak at the same mass. This ensures that both the {\it Spitzer} and {\it Chandra} surveys have the same completeness. There is a large population of low mass cluster candidates, down to 0.4 ~M$_{\odot}$ which lack a X-ray counterpart.  % and the mass limits ???  (as done in ???METTIAMO QUALCHE REFERENZA). \\
%The mass function is computed using the 4~Myr and the 5~Myr isochrone to
%convert the 3.6$\mu$m magnitudes to stellar masses (see Fig.~\ref{massfun}). 
%The observed mass function departs from the theoretical IMF \citep{Kroupa2002} at M$\sim$1.0\,M$_{\odot}$ 
The observed mass function falls below the theoretical IMF below $\sim 1.0$~M$_\odot$. % [1.2\,M$_{\odot}$]
% using an isochrone of 4~Myr [5~Myr]. %,  to derived mass function from the observed magnitudes.
 This mass corresponds to the estimated completeness at 3.6$\mu$m of 14.0~mag. \\
Extrapolating the total population of IC~1795 down to a mass of 0.08~M$_{\odot}$, the total cluster membership numbers $\sim$2000 stars. 
This corresponds to a stellar density of $\sim$15 stars $pc^{-3}$, which is several thousands lower than the density in the Trapezium cluster. \\
In summary:
\begin{itemize}
\item IC~1795 has an age between 3 and 5~Myr consistent with the previous estimate of \citet[][]{Oeyetal2005}, 
\item the mass function is well-reproduced by a standard IMF to a mass limit
of 1~M$_{\odot}$-1.2~M$_{\odot}$,
\item the cluster population is estimated to be 2000 sources at a mass limit of 0.08~M$_{\odot}$.
\end{itemize}
\section{Disk evolution: effect of the stellar mass and the environment}
\label{discussion}
In this section the analysis of the disk fraction as a function of the stellar mass and distance from the center of the cluster is presented for the cluster members\footnote{defined in section~\ref{membership} as infrared sources with an X-ray counterpart.} of IC~1795.
%In this section, we analyze the disk-bearing cluster members\footnote{defined in section~\ref{membershipX} as {\it Spitzer} sources with an X-ray counterpart}, as function of the (1) stellar mass, (2) spatial distribution of cluster members, (3) distance from the O-type star. 
%Finally we discuss our results in the context of previous studies of disk evolution
%in low- and high-mass environments.
\subsection{Disk fraction vs stellar mass and spatial distribution}
\label{diskfrac-mass}
\begin{figure*}
 \centering
  \includegraphics[width=13cm]{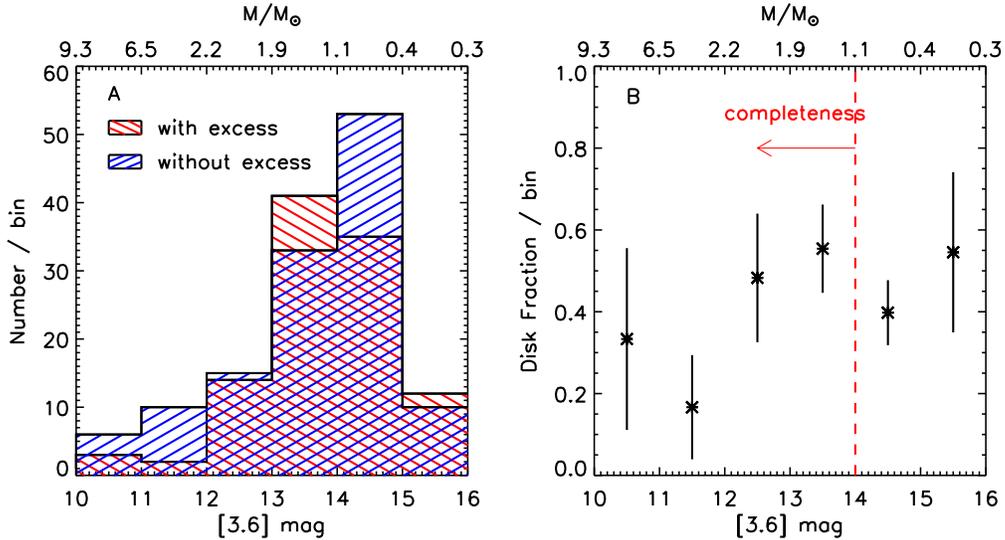}
  \caption{
    {\it A}: Histogram of cluster members (e.g. IRAC sources with an X ray
      counterpart) sources with excess (i.e. with
    [3.6]-[4.5]$>$0.2) and without excess with a bin of 1 magnitude.
     {\it B}: Disk fraction computed as ratio of the
     sources with  excess per [3.6] bin and the total number of sources per
     bin. The corresponding masses are computed using the 4~Myr isochrone.
%      {\it A-B}: all IRAC sources; {\it C-D}: IRAC sources with an X ray
   %   counterpart; {\it E-F}: IRAC sources without an X ray counterpart.}  
}	
  \label{df-irac}
\end{figure*}

\begin{figure}
 \centering
  \includegraphics[width=13cm]{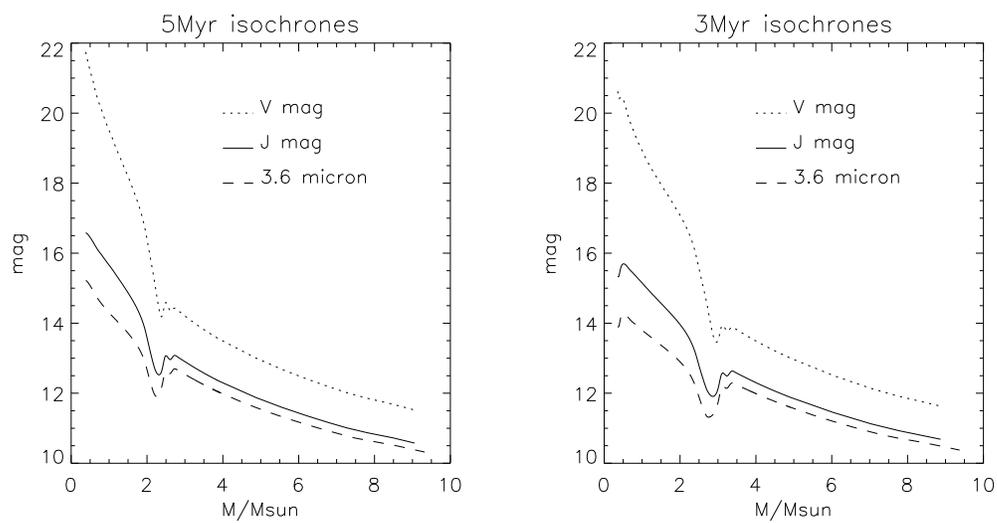}
   \caption{Relation between the V, J and 3.6$\mu$m magnitudes and the stellar
     masses computed by the 5~Myr ({\it left}) and 3~Myr ({\it right})
     isochrones. 
            }
      \label{iso_info}
 \end{figure}

\begin{figure}
 \centering
  \includegraphics[width=6.5cm]{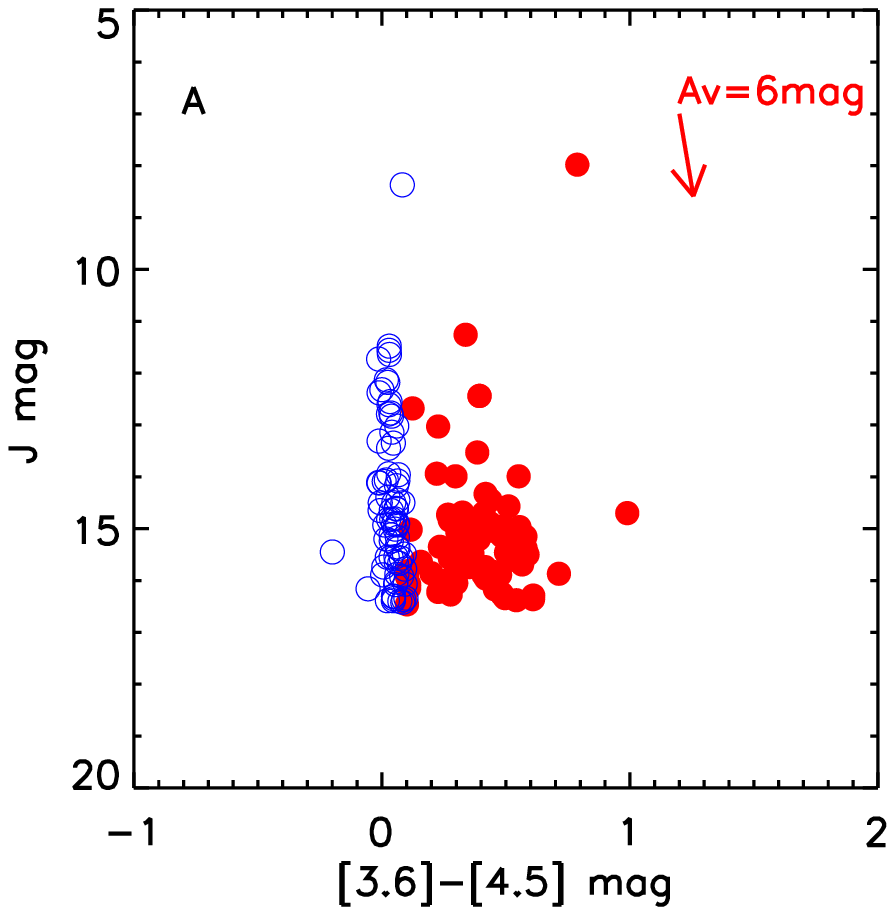}
  \includegraphics[width=13.cm]{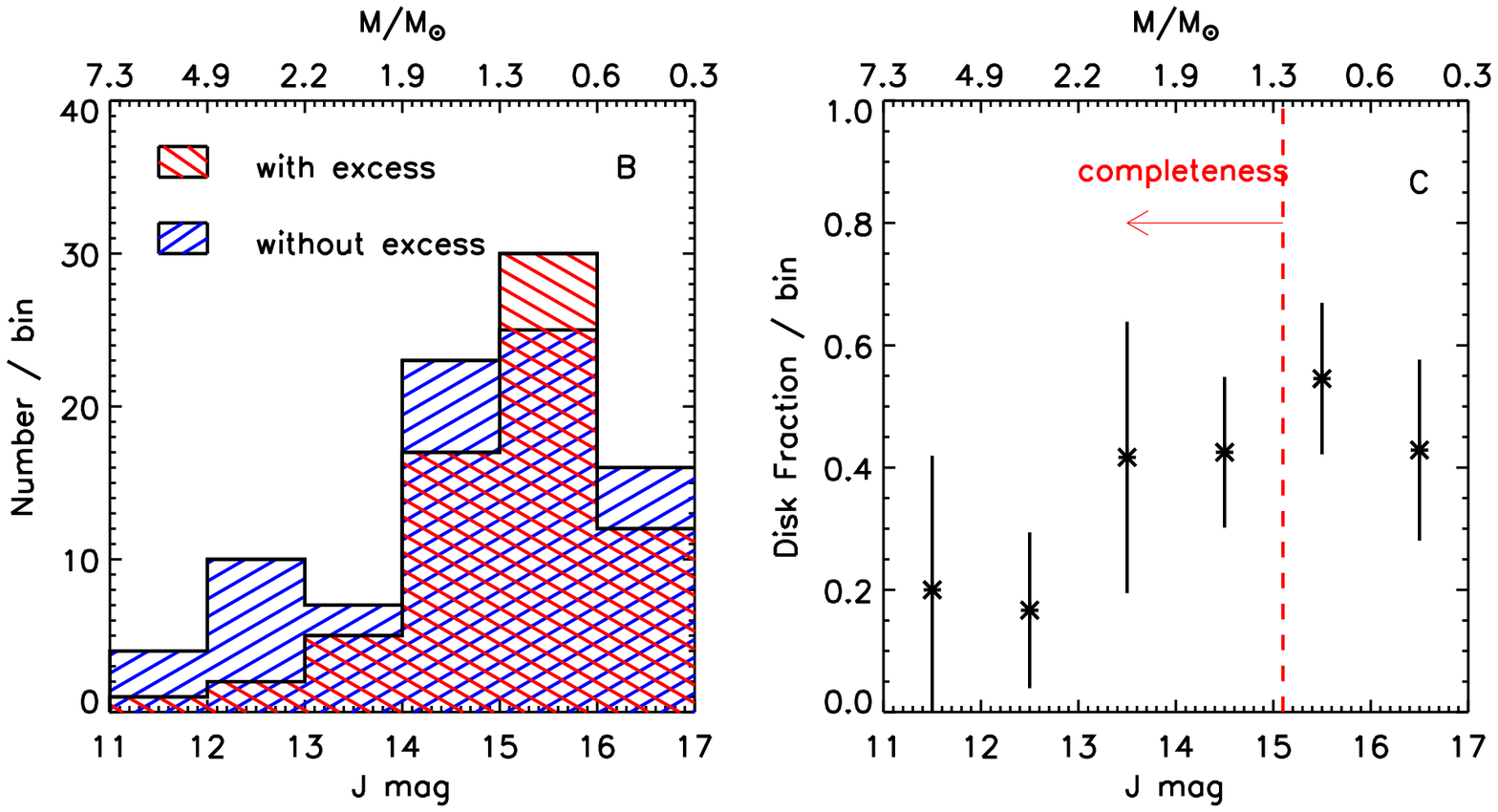}
  \caption{
    {\it A}: J -- [3.6]-[4.5] CMD. The arrow represents an
    extinction A$_V$=~6~mag.  Filled circles represent sources with excess (i.e. with [3.6]-[4.5]$>$0.2), and empty circles represent sources without excess. {\it B}: Number of sources with and without excess per bin of J=1~mag.
     {\it C}: Disk fraction computed as ratio of the sources with
     excess per J~mag bin and the total number of sources per bin. The corresponding masses are computed using the 4~Myr isochrone.
  }
  \label{histJ}
\end{figure}

\begin{figure}
  \centering
  \includegraphics[width=12cm]{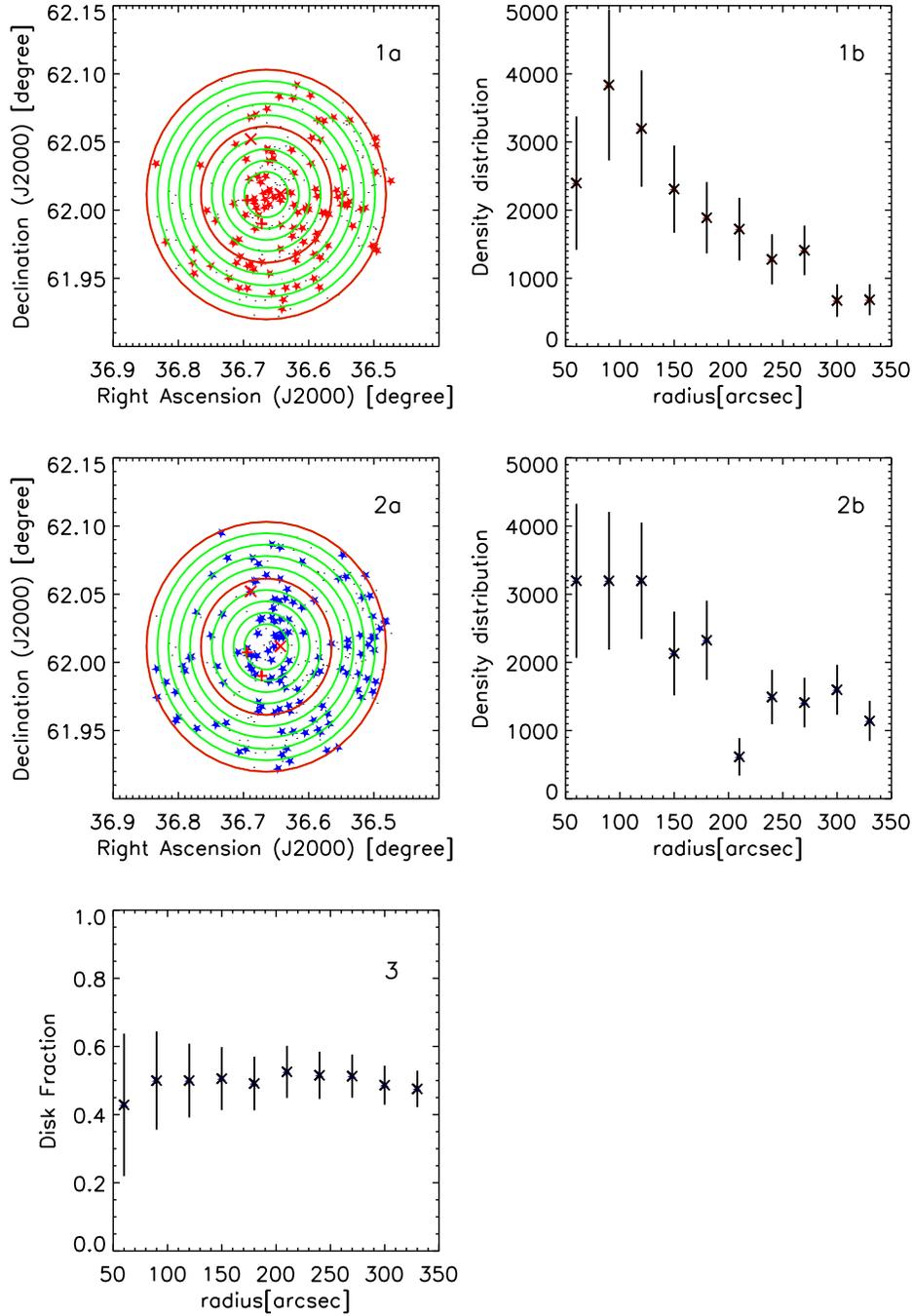}
    \caption{Spatial distribution ({\it 1a \& 2a}) and spatial density distribution
    ({\it 1b \& 2b}) of the sources with ({\it 1a \&1b}) and without ({\it 2a \& 2b})
    infrared excess. The excess has been selected with [3.6-4.5] $>$ 0.2 from
    the panel B of Fig.~\ref{cc+am_all}. The circle are as in
    Fig.~\ref{distr_all}.  The '$\times$' and '+' symbols represent the positions of the O and B stars in the cluster. In panel {\it 3} the disk fraction is computed as a
    function of the distance from the center of the cluster.
      }  
  \label{exc-noexc-distr}
\end{figure}

%Fig.~\ref{df1}-\ref{df2}
 We examine the mass dependence of disk emission 
  of the 3 - 5~Myr old cluster IC~1795. 
 Since only few stars have been spectrally classified so far \citep{Oeyetal2005}, 
 and a detailed spectral classification of
 the cluster members is in preparation (Kim et al.~2010, in preparation), 
% the stellar masses are computed using the mid-infrared isochrones (details in 
% section~\ref{secmassfunc}).\\
 the disk fraction is computed using the 3.6$\mu$m and the {\it J} magnitudes as a
 tracer of stellar photosphere (Fig.~\ref{df-irac} and~\ref{histJ}, details in section~\ref{secmassfunc}). \\
%Since this is the first paper where the disk fraction is computed using the IRAC magnitudes and derived from them 
% the corresponding stellar mass, 
In order to calculate the fraction of sources with and without infrared excess, we follow the method described in Appendix~\ref{diskfrmidir}. 
   We notice that, between $[3.6]=10-12$, the disk fraction rapidly increases
   from 20\% to 50\%, and then it remains constant at 60-40\% between 12 and 14.5~mag. 
   Constructing a $2 \times 2$ contingency table of IR excess for the $[3.6]=10-12$ and $12-14$ magnitude ranges, we formulate a hypothesis test with null hypothesis that IR excess does not depend on $[3.6]$ or, equivalently, on stellar mass. A Fisher's exact test applied to the contingency table indicates the probability of this null hypothesis is $P=0.005$. The test is implemented by the {\em R} statistical software package \citep{R10}. We conclude that the effect is statistically reliable. 
%   This value was found also in the analysis of the disk fraction as a
 %  function of the spatial distribution of the cluster members. The reason for having the same value in the last two analysis is that the almost flat shape of the mid-infrared isochrones (see  Appendix~\ref{iso}) does not allow a direct conversion betweens a magnitude and a mass: this means that the disk fraction at 3.6$\mu$m~= 14-15~mag is mixing stars of different masses (between 1.1 and 0.38~M$_{\odot}$).\\ 
Since $\sim$80\% of our IRAC sources have a near infrared counterpart, 
we used the {\it J} magnitude as a proxy for stellar mass \citep[see, e.g.][]{Hernandezetal2007a}. These bands have the advantage of being relatively unaffected by disk excess emission. However, they are strongly affected by extinction. 
The 4 Myr isochrone (see section~\ref{age}) has been used to infer masses from the {\it J} magnitudes. In this band the relation between mass and magnitude is almost unique (see Fig.~\ref{iso_info}).\\  
   In Fig.~\ref{histJ} (panel A) we show the CMD using the {\it J} magnitude as a photospheric
   tracer and the [3.6]-[4.5] IRAC colors to define the sources with and
   without infrared excess. After the histograms of sources with and without excess
   (panel B of Fig.~\ref{histJ}), the disk fraction is shown in panel
   C. This is computed as the ratio of sources with and without
   excess in bins of one magnitude. The errors are computed as the square
   root of the number of sources in each bin. The corresponding mass is
   reported in the upper x-axis.\\ 
   The disk fraction is found to increase from 20\% to 60\% for {\it J} magnitudes 
   from 11 to 15. This corresponds approximately to masses between 8.4 and
   0.8~M$_\odot$. The disk fraction remains constant around  50\% toward lower
   masses.   An increasing frequency of disks is obvious down to 15 magnitudes even after taking into
   account the uncertainty in the disk fraction. \\
%   The decreasing disk fraction
%   for masses smaller than 0.8~M$_\odot$ might be due to the incompleteness  
%    of the survey below 0.8~M$_\odot$.\\}
   Taking into account solar-type stars only with masses $\sim$1~M$_\odot$ (our completeness level,  section~\ref{secmassfunc}), the disk fraction is 50\%.\\ 
This value is consistent with the disk fraction of solar-type stars previously computed with the 3.6$\mu$m magnitude. 
%   In Fig.~\ref{histV} we investigate the changes in disk fraction as a function
 %  of stellar mass, using as photospheric tracer the {\it V} magnitude, instead
%   of {\it J}. Also in this case, the disk fraction increases with decreasing {\it V} magnitude.\\
A dependency of the disk fraction on the stellar mass was found in {\it Spitzer} surveys of 
young clusters with low and high-mass members \citep[e.g.][]{Hernandezetal2007a, Carpenteretal2006, Currieetal2007, Ladaetal2006, Luhmanetal2010}. The general finding is that massive stars seem to loose their disk earlier than lower mass stars. 
This was also found in clusters older than 3~Myr by \citet{KennedyKenyon2009}, using published optical spectra and infrared excess data.\\
 %n analysis of the disk fraction as a function of stellar mass was presented by \citet{Hernandezetal2007a} in a {\it Spitzer} study of $\sigma$Ori. In their work the lowest fraction of disks was observed in the highest mass stars (in the HAeBe range). They found a marginal evidence that the disk fraction declines toward lower masses (into the brown dwarf range), in agreement with results from \citet{ladaetal2006} for the young  (2-3 Myr) stellar group IC~348. Nevertheless, including the error bars, the disk fraction in $\sigma$Ori was also consistent with no mass dependence toward lower masses. \\
%\subsection{Disk fraction vs spatial distribution}
In young clusters the disk fraction may depend also on the spatial position. % when e.g. star formation was triggered. 
If triggered star-formation took place in IC~1795 itself, a different spatial distribution 
of sources with and without disks and with an age gradient is expected \citep[e.g. Cepheus~B, Tr~37, ][]{Getmanetal2009, Sicilia-Aguilaretal2005, Sicilia-Aguilaretal2006a}. 
%Previous studies proposed that the different star-forming regions in the W3 
%molecular-cloud have been triggered by ionizing winds from IC~1795 \citep[][]{Oeyetal2005, FeigelsonTownsley2008}. }
 In Fig.~\ref{exc-noexc-distr} we show the spatial density distribution of
 sources with and without infrared excess. \\
 The disk fraction ($f_d$) is defined as the ratio between the number of sources with
 disks ({\it N$_{\rm disk}$}) and the total number of sources (with and without
 disks; {\it N$_{\rm tot}$}). The fraction $f_d$ was computed in consecutive rings.
 Within the uncertainties, the disk fraction remains constant across the
 IC~1795 region at about 50\%.\\ 
 This result argues against a 
triggered star formation scenario for IC1795.
A constant disk fraction as a function of the distance from the center of the
cluster was also found e.g. in $\sigma$~Ori \citep{Oliveiraetal2006} using {\it K} and
{\it L}-band observations. This result is however in contrast with
\citet{Hernandezetal2007a} who instead found an evidence for a higher disk
fraction near the cluster center for $\sigma$~Ori. 
 
\subsection{Influence of the O star in the cluster: disk photoevaporation and cluster
dynamics}
\begin{figure*}
  \centering
\includegraphics[width=20cm,  trim=6 29 5 10, clip, angle=90]{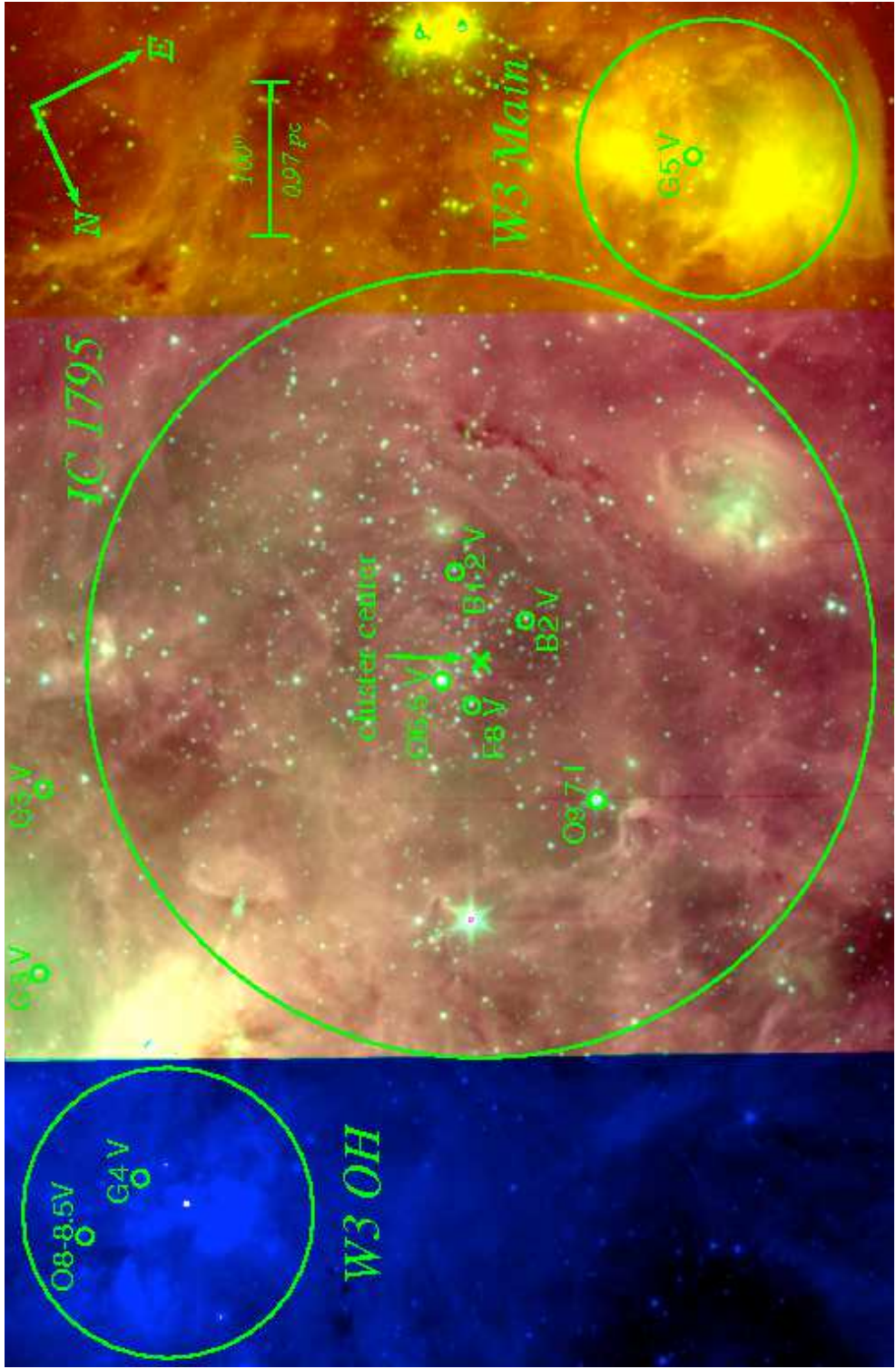}
  \caption{Combined image of the mosaics obtained with the long exposures of IRAC/ch1, IRAC/ch2 and IRAC/ch4. Highlighted are the edge and center of IC~1795 together with the positions of the high- and intermediate-mass stars in the region spectrally classified by \citet{Oeyetal2005}. W3-Main and W3-OH are not covered by all IRAC channels.
  }      
  \label{prova}
\end{figure*}

%In this section we analyze the effect of the O9 star on the disk fraction and dynamics of IC~1795.\\
In Fig.~\ref{prova}, we notice that the O6.5V and the O9.7I stars lie at 39\arcsec\,west and 120\arcsec\,north respectively, from the projected spatial center of the cluster defined in sec.~\ref{IRACpositions}.
In order to study the influence of the O stars in the cluster we have re-centered the cluster on each O star. 
In both cases, the spatial distribution of the disk fraction observed is almost constant  with distances from 
 each O star including within the first 60\arcsec\, (which corresponds to 0.58~pc at the cluster distance). \\
 This result is not in agreement with previous works of \citet{Balogetal2007} and \cite{Merceretal2009}. From the analysis of
the 2-3~Myr NGC~2244 cluster \citet{Balogetal2007} found a smaller disk
fraction within 0.5~pc from the O stars, including the two highest-mass O stars, HD 46223 (O5) and HD 46150 (O6) of the cluster.\\
\citet[][]{Johnstoneetal1998} predicted an absence of disks in systems proximate to O-stars, a prediction supported observationally by Hernandez et al. (2008) who found no evidence for primordial disks within 0.75 pc of an high-mass binary (consisting of an O 7.5 star and a Wolf-Rayet star) in the $\gamma$\,Velorum star-forming region.\\
However,  theoretical expectations predict the photoevaporation of the outer part 
of the disk ($>$ 5 AU), while in the inner regions (traced by {\it Spitzer/IRAC}), where the escape velocity exceeds the sound speed of the ionized gas, cannot be evaporated. The dissipation of such a part of the disk proceeds via viscous transport of material from the inner to the outer disk  \citep[e.g.][]{Adamsetal2004}. 
Models predict that a high-mass star introduces an external 
UV radiation which can photoevaporated  disks only within 0.3-0.7~pc \citep[e.g.][]{Johnstoneetal1998, Adamsetal2004, Clarke2007, Gortietal2009}. 
 % compared to the fraction obtained at
%larger distances. 
Finally, we investigate whether mass segregation acted on the dynamics of the cluster. \\
%Although some degree of mass segregation occurs earlier, the position of massive 
%stars in rich young clusters generally reflects the cluster's initial conditions. 
%{\bf However, mass segregation moves massive stars in the center of young clusters. 
%In the Trapezium cluster in Orion e.g.,  \citet{Bonnelletal1998} suggested that the position of the high-mass stars indicates that they formed in, or near, the center of the cluster.\\ 
In IC~1795 the high-mass stars are not in the center of the cluster, nor are they preferentially concentrated in a particular region compared to lower mass members (see Fig.~\ref{prova}). The exception is represented by the O6.5V star which is only 38\arcsec\, away from cluster center. Since the spatial distribution of high-mass stars does not allow to conclude if mass segregation has taken place in IC 1795, we check whether the cluster is in dynamical relaxation. 
%This is a first evidence that no mass segregations has taken place in IC~1795. 
The time scale of mass segregation can be approximated  by the relaxation time ($\mbox{$t_{\rm relax}$ }$
\footnote{
$t_{\mathrm{relax}\,}\approx ( N/ 8\mathrm{ln}\,N) t_{\mathrm{cross}}\,$, where $t_{\mathrm{cross}\,}=2R/ v_{\mathrm{disp}\,} $ is the crossing  timescale and $v_{\mathrm{disp}\,}$ the velocity dispersion, \citep[][]{Bonattoetal2006}.
) in which a cluster reaches some level of kinetic energy equipartition, with the massive stars sinking to the core and low-mass stars moving to the cluster halo \citep[e.g.][]{Bonattoetal2006}.  }). 
%We estimate the two-body dynamical relaxation time scale $t_{\mathrm{relax}}$
%of IC~1795 using the formula \citep[e.g.][]{Bonattoetal2006}%:  
%\begin{equation}
%t_{\mathrm{relax}\,}\approx ( N/ 8\mathrm{ln}\,N) t_{\mathrm{cross}\,} 
%\end{equation}
%where $t_{\mathrm{cross}\,}=2R/ v_{\mathrm{disp}\,} $ is the characteristic time
%for a star to cross a cluster of radius $R$ and velocity dispersion $v_{\mathrm{disp}\,}$.  
Adopting a radius of the cluster $R\sim3.2pc$ (see section~\ref{IRACpositions}) and using a
rough estimate for the unmeasured velocity dispersion $v_{\mathrm{disp}\,}\sim
3  km s^{-1}$ \citep[][]{BinneyTremaine1987}, and $N\sim2000$  stars (see section
\ref{secmassfunc}), we obtain $t_{\mathrm{relax}\,}\sim$~68~Myr for IC~1795. 
%We note that the actual number of cluster members may be higher than described here, 
%leading to an even larger $t_{\mathrm{relax}}$. \\
As the cluster age is $\sim$17 times less than its relaxation time, 
this calculation supports the observational evidence that mass segregation has not occurred in IC 1795. A similar result has been found in the 2-3 Myr old cluster NGC~2244
by \cite{Wangetal2008,Wangetal2009}. However, in young rich clusters, massive stars are usually concentrated in the center %while lower-mass cluster members at larger radii from the center 
\citep[e.g. ONC, ][]{Hillenbrandetal1998}. 
% Another possible explanation of the absence of mass segregation is that cluster members are not all coeval and that the massive stars have been scattered into the cluster \citep[e.g. M17, ][]{Nielbocketal2001}.

\subsection{Disk evolution in low and high-mass environments} 
\label{diskev}
\begin{figure}
  \centering
   \includegraphics[width=14cm]{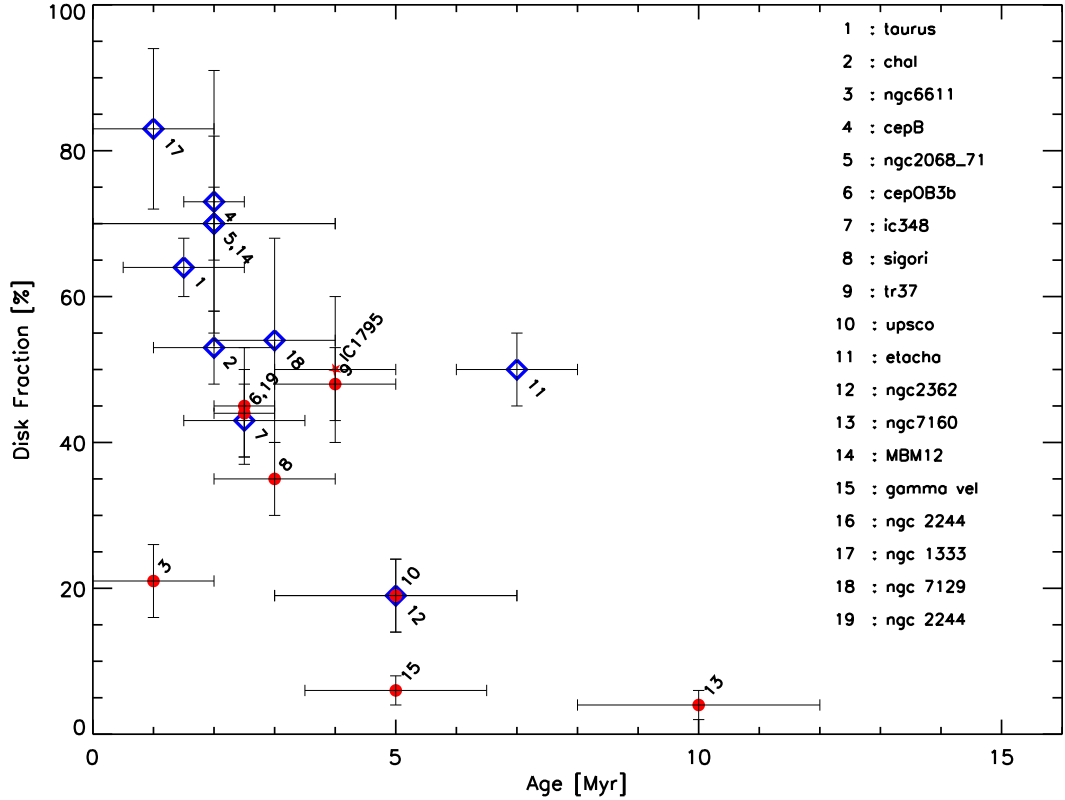}
    \caption{Disk fractions computed in the IRAC colors, as a function of the age of the cluster. The filled dots represent the OB associations, while the empty diamonds represent the low-mass star-forming regions. }
       \label{evol}
  \end{figure}

In this section the disk fraction of IC~1795 is compared with the results obtained in other low and high-mass star-forming regions with the aim of investigating the influence of high-mass stars on the disk dissipation timescale. 
In particular, using the mid-infrared wavelength range, which traces the circumstellar dust at a few AU from the star, it is possible to constrain the dissipation timescale of the region of the circumstellar disks where planets form.\\
The disk fractions computed using the IRAC/{\it Spitzer} colors have been compiled from the literature and compared with the result obtained on IC~1795. 
%In IRAC colors, 50\%$\pm$10\% of cluster members of IC~1795 with masses $\sim$1~M$_{\odot}$  are found to have a disk (details in section~\ref{diskfrac-mass}).\\
The clusters considered are listed in Table~\ref{diskfrac}. The total disk fraction from the individual studies  is presented highlighting the 
completeness  mass range of their work. %with the mass range where their work was complete. 
When available, the disk fraction only in the intermediate-high-mass range is also reported. 
%We used in particular the results of the binning in mass done on a sample of clusters collected from the literature from \citet{KennedyKenyon2009}.\\
%For these reasons we plan to extend our survey to younger (1 - 5 Myr) OB
%associations to consistently investigate the role of the external UV radiation
%on the evolution of circumstellar disks around solar-type stars.
%- the 1~Myr Taurus and the Coronet cluster at \citep{Luhmanetal2010, Sicilia-Aguilaretal2008}, \\
%- IC~348 at 2-3 Myr \citep{ladaetal2006}, \\
%- $\sigma$~Ori at 3 Myr \citep{Hernandezetal2007a},\\  
%- Tr 37 at 4 Myr \citep{Sicilia-Aguilaretal2006a, Merceretal2009}, \\
%- NGC 2362 at 5 Myr \citep{DahmHillenbrand2007}. \\
When the errors on the disk fractions ($\Delta f$) are not available in the literature, we used $\Delta f = \frac{1}{\sqrt[]{N_{\rm TOT}}} $, where $N_{\rm TOT}$ is the total number of cluster members.\\
%In particular, in the case of MBM~12 the disk fraction is computed as follows: \citet{Meeusetal2009} detected 8 of 12 cluster members with MIPS; of these 8, 7 show an infrared excess. Considering that additional 4 sources not detected by MIPS so far may have all an infrared excess, we consider that the disk fraction may vary between 58\% (7/12) and 83\% (10/12) .\\
%The percentage of members with disks in NGC~6611 is computed as the weighted mean of the disk fractions computed at different incident fluxes from the OB members in the cluster, which vary between 31\% and 16\% \citep[][]{Guarcelloetal2009}.\\
%However, since the cluster membership has been defined, in most of the cases, using optical spectroscopy (as we discussed in section~\ref{membership}), the cluster memberships are not complete for intermediate mass objects.\\
The general trend of disk dissipation in low-mass and high mass environments (see Fig.~\ref{evol}) suggests that in young clusters (1 Myr) the fraction of disks is about 90\%-80\%. By about 3-4 Myr the fraction of disks is reduced to 50\%-40\%. At 5 Myr the disk fraction drops to 20\% while after 10 Myr almost all disks are dissipated. \\
Apart from the general trend previously described, the disk fraction of clusters containing more than 5 O/B stars shows mass dependent effects. For the O and B stars, the disk fraction decreases to 30\% in the first 3 Myr, while in the low-mass environment it is still 60\%-85\%.\\
%Apart from the general trend previously described, the disk fraction 
%of clusters with more than 5 O/B stars drops already down to 30\% 
%in the first 3 Myr, while in low-mass environment it is still 60\%-85\%. \\
%The disk fraction of $\gamma$Vel lays significantly below the general trend 
%for a cluster at 6~Myr. 
%However, if this fraction of disk 
%Without distinguish in spectral type nor to the different sensitivity of the cluster surveys compiled from the literature, the fractions of cluster members with disks in these clusters decrease from 80\% to ~20\% with age (Fig.~\ref{evol}). \\
Our result is consistent with the disk fraction found in cluster Tr 
37 \citep{Sicilia-Aguilaretal2006a}, an OB association with an age similar to that of IC~1795.\\
Most of the disk fractions from the literature are computed for clusters within 1kpc. %This allows them to have a membership down to M-type stars.  \\
IC~1795 is at $\sim$2~kpc and our study is complete to $\sim$1~M$_\sun$, which corresponds to early-K spectral types. Therefore a comparison between the disk fraction of 50\%$\pm$10\% found for IC 1795 and that of other clusters might be misleading, e.g. for lower masses where our survey is incomplete, we expect to miss a significant fraction of the photospheric population.\\
A number of bias effects in compiling disk fractions from the literature, (e.g. the completeness of the cluster memberships, the 
uncertainty on the age and the disk fractions not specified in different mass range), %might complicate the disk fractions compiled from literature and
have been carefully highlighted, since they can affect our conclusions. 
%In particular,  may and the disk fractions which is not always computed for a specific stellar mass
%range;\\ 
%- unresolved binarity.\\
Given all uncertainties, we find that IC~1795 follows the smooth
decline of the disk fraction with age observed for other clusters. 
%This suggests that only the presence of several high-mass stars in a cluster 
%is needed to accelerate the dust dissipation in disks.\\ 
\section{Summary}
\label{conclusions}
In this paper we presented a deep {\it Spitzer}/IRAC survey of the OB association
IC~1795.\\% to investigate the influence of external UV radiation on
%protoplanetary disk evolution. \\
Combining the infrared {\it Spitzer}/IRAC observations with a deep X-ray
{\it Chandra}/ACIS survey we carefully establish the cluster membership of the
cluster. 
Compared to previous optical based studies on young clusters, this is the only method
which allows for a determination of the disk fraction for cluster members with 
masses $\ga 1.0$~M$_\odot$. %G-type stars.
The IR/X-ray surveys are complete down to 1~M$_{\odot}$ or 1.2~M$_{\odot}$
computed with the 4~Myr or 5~Myr isochrones, respectively. The age of the cluster 
is determined via the position of the Class III stars (stars with no disks) on the 
[3.6] -- [3.6]-[4.5] color-magnitude diagram. 
%Compared to optical and near-infrared CMDs, this diagram has the advantage that
%different isochrones are more separated and the interstellar extinction is
%negligible. 
For IC~1795 we determine an age of 3 -- 5 Myr.\\
%IRAC sources with infrared excess (characteristic of circumstellar disk
%around young star) without X-ray counterpart are classified as cluster
%member candidates.\\
The spatial distribution of the cluster was found to be asymmetric
around the cluster center and no mass segregation is present.\\%, nor around the massive O and B type stars.\\
The disk fraction was analyzed as a function of the distance from the
cluster center as well as a function of the stellar mass. 
%In the first case, the fraction of disk was computed in a series of
%concentric circles (each 30\arcsec) from the center to the outer radius: within
%the errors, t
The objects with disks represent the 50\% of the total source number.  
No spatial dependence of the disk frequency was found.\\
%This suggests that only one star-formation event occurred in IC~1795 and 
%that probably it was not triggered by external ionization shocks.\\
Using the [3.6] IRAC magnitude as a tracer for the stellar mass we find
that the disk fraction is $\sim$20\% for masses $>$2~M$_{\odot}$ and 
 $\sim$50\% for masses $<$2~M$_{\odot}$. %, down to our survey completeness:  
 We confirm that the dissipation of disks around high-mass stars 
 ($>$2~M$_{\odot}$) is faster compared to the dissipation around  stars of 1-2~M$_{\odot}$.\\
%increases from 20\% to 50\% -- 60\% as the stellar mass
%decreases from 8 to 2~M$_\odot$ and remains constant at 50\% in the mass
%range 0.4 -- 2M$_\odot$. 
%We note that in the case of solar mass stars
%($\sim$1~M$_\odot$) we find a disk fraction of 40\% (instead of 50\%)
%when using the J magnitude as tracer for the stellar mass.\\
%The presence/absence of disk in the vicinity of the O-type star has been
%analyzed to search for any dependency of disk evolution from external UV
%radiation. 
 We found no variation in the disk fraction within 0.6 pc of the O-type stars in the association.\\
%the single O-type star in the OB association.\\
%The study of the evolution of disks in different environment is necessary to
%constrain the timescale of the formation of planets.
%Further investigation is needed to decipher in particular the role of external
%UV radiation on disk lifetime.  
%Apart of NGC~2244, is the only cluster where 
%No evidence of mass segregation has been found in IC~1795.\\
Measurements of the disk fractions in low-mass and high-mass environments have been collected from the literature. We found that, in general, the disk dissipation timescale is comparable in high-mass and low-mass clusters.
%Measurements of the disk fractions in low-mass and high-mass environment have
%been collected from the literature and we found that in general the temporal evolution 
%is comparable in high-mass and low-mass clusters. %Only the presence of more than five 
%high-mass stars in a cluster seems to accelerate the disk dispersion.
\acknowledgments

We are grateful to the {\it Spitzer Science Center} for their support during
the data reduction. We acknowledge Leisa Townsley for supplying the Chandra/ACIS sourcelist for W3 and thank Patrick Broos for assistance with matching those X-ray sources to the IRAC sources. We acknowledge %Jessy Jose for providing the optical photometry of the cluster and 
G. Rodighiero for the helpful discussion about the extragalactic contamination in the mid-infrared.  
This publication makes use of data products from the Two Micron All Sky Survey, which is a joint project 
of the University of Massachusetts and the Infrared Processing and Analysis Center/California Institute 
of Technology, funded by the National Aeronautics and Space Administration and the National Science Foundation. 
E.D.F. was supported by Chandra grant SV4-
74018, NASA grant NNX09AC74G and NSF grant AST-0908038. ASA is supported by the
Deutsche Forschungsgemeinschaft (DFG) grant number SI 1486/1-1.

\appendix
\section{Evolutionary model}
\label{iso}
The isochrones adopted for the analysis have been computed using the FRANEC
evolutionary code. 
We briefly describe here the physical inputs in the code. 
For detailed explanation the reader can refer to  
\citet{Chieffietal1989,Deglinnocentietal2008}. The 
opacity tables are from \citet{Fergusonetal2005} for $\log T[K] < 4.5 $ and
from \citet{Iglesiasetal1996} for higher temperatures. 
The equation of state (EOS) is described in \citet{Rogersetal1996}. Both
opacity tables and EOS are calculated for a heavy elements mixture equal to
the solar mixture of \citet{Asplundetal2005}. Our models are completely
self-consistent, with a {\it unique} solar chemical composition, $(Y, Z) =
(0.27, 0.02)$. The value of the mixing length
parameter adopted in the models is $\alpha_{MLT}= 1.6$.\\
Transformations from the theoretical $(log(T_{eff} [K])$, $log(L/L_{\odot}))$
to the observational planes have been performed via synthetic photometry.  In
particular we have computed the isochrones in the 2MASS and $VI_{\rm Bessell}$
photometric systems and for the first time in the IRAC photometric system. 
 The optical and near-IR isochrones are consistent with the PMS isochrones 
computed by \citet{Siessetal2000}.
The filter throughputs and zero points of the IRAC photometric system are defined
in \citet{Reachetal2005}. We used stellar spectra generated with both the ATLAS9 (see
\citet{2003IAUS}) and PHOENIX (\citet{Brottetal2005}) model atmospheres
codes. 
 
The ATLAS9 grid of models is limited to temperatures higher than 3500 K;
corresponding isochrones have then a lower cut in mass at values of 0.36
M$_\odot$ and 0.38 M$_\odot$ for 3 and 5 Myr respectively. 
The PHOENIX grid can cover a region of lower temperatures (T $> 2000$ K) but
has an upper limit in temperature of 10000 K. This corresponds to 2.96
M$_\odot$ and 2.56 M$_\odot$ for 3 and 5 Myr respectively. 

%Since we used the approximation of gray atmosphere these isochrones cannot
%reproduce stellar atmospheres colder than ???, which correspond to a stellar
%mass of 0.38~M$_{\odot}$.
\subsection{Mid-infrared isochrones}
In the mid-infrared (MIR) CMDs, the part of the isochrone for
intermediate-to-high-mass stars that have reached already the Main Sequence (MS) is almost vertical.
Lower mass stars are still in the PMS phase. The PMS objects
are cooler and with lower surface gravities; their spectra at the MIR
wavelengths are no longer described by an (approximate) black body
exponential tail, as in the MS phase.
Due to the presence of spectral features the PMS isochrones show tendency towards redder colors. 
\\
%What is very interesting is the ``evolution of the overall shape'' of the
%isochrones in the PMS phase with time.
PMS stars have larger radii when they are at the top of their Hayashi
track, and the radii become smaller as the stars evolve and contract. In the
totally-convective phase the effective temperature changes little,
leaving the observed color almost unchanged, even though the change in
surface gravity slightly affects the shape of the emerging spectrum. 
But with decreasing age, the change in radius heavily affects the total
luminosity.
For this reason mid-infrared PMS isochrones are no longer age-degenerate as they are in the MS phase. On the contrary different PMS isochrones are well separated in the MIR CMDs and can be used for age-dating the cluster, or, at least, setting some constraint on its age. 
In particular, the transition phase between PMS to MS is characterized by a hook-shaped bending in the isochrone. 
As the age increases, the hook and the red PMS branch move
 towards larger magnitudes (lower luminosities). 
When an independent estimate of the distance and the extinction is available, as in our case, the Main-Sequence Turn-On (MSTON) can be used as upper limit for the age of a cluster. After placing an isochrone in the CMD, if no star is observed below the MSTON, at color equal zero, then the cluster must be younger than the given isochrone.\\
For this reason the position of the MSTON is used to infer the age of the cluster.
%% \begin{figure*}
%%  \centering
%%   \includegraphics[width=15cm]{match_20may/iso_info.ps}
%%    \caption{Relation between the V, J and 3.6$\mu$m magnitudes and the stellar
%%      masses computes by the 5~Myr ({\it left}) and 3~Myr ({\it right})
%%      isochrones. 
%%             }
%%       \label{iso_info}
%%  \end{figure*}

\section{Disk fraction in the [[3.6],[3.6]-[4.5]] CMD}
\label{diskfrmidir}
\begin{figure}
  \centering
   \includegraphics[width=14cm]{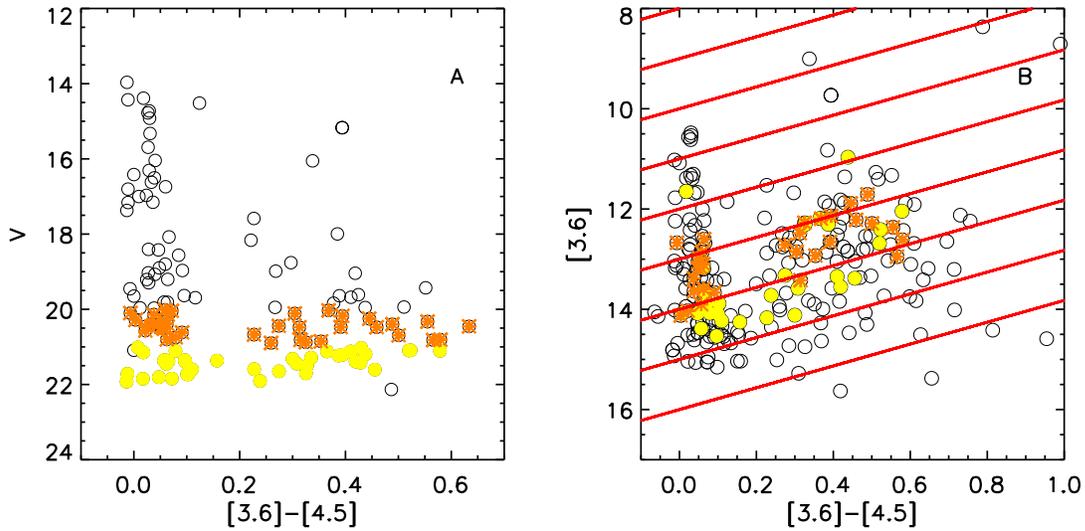}
       \caption{{\it A: }[[3.6]-[4.5], V] CMD. Highlighted with different
      symbols are the stars in 2 consecutive horizontal slices with V between 20
      and 21 and with V between 21 and 22. {\it B:} [[3.6]-[4.5],[3.6]] CMD.
      Highlighted are the same sources as in panel A. The derived lines used
      to compute the disk fraction are overplotted.
             }
       \label{dfrmir}
 \end{figure}

Fig.~\ref{dfrmir}A shows the [[3.6]-[4.5], V] CMD. 
%Since V magnitude is directly proportional to the stellar mass.
In this diagram the presence of the circumstellar disk causes the object to
move horizontally towards red colors. This is because the disk emission
affects only slightly the V magnitude (e.g. through optical veiling produced by
mass accretion). \\
 In the [[3.6]-[4.5],[3.6]] CMD shown in Fig.~\ref{dfrmir}B, disk emission also affects the stellar luminosity ([3.6] mag). As a consequence the presence of the disk will move the
object towards red colors obliquely.\\ 
To properly compute the disk fraction as a function of the [3.6] magnitude
(and hence stellar mass) it is necessary to take into account this effect. 
We calculate the isomass in the [3.6]
magnitude in the [[3.6]-[4.5],[3.6]] CMD. This was computed in the following
way: 
we select all stars in a series of horizontal slices in the [[3.6]-[4.5], V]
CMD (central V = 20.5; 21.5 mag and bin = 0.5 mag). We look for the position
of the same objects (highlighted with the same symbols) in the
[[3.6],[3.6]-[4.5]] CMD. We extrapolate the median inclination of the position of these
sources in this CMD for each magnitude bin.
Finally, we computed the disk fraction in a series of bins counting the
sources in the oblique slices.

\section{Relation between 3.6$\mu$m IRAC magnitude and stellar mass}
\label{mag-mass}
Due to the spread of the sources along the isochrone, we assign the most
probable mass by taking into account the photometric uncertainty via a
maximum-likelihood method. We define the Likelihood function for the i-th
observed star as: 
\begin{equation}
\label{eq:lkl}
\mathcal{L}^i(m_j) = \frac{1}{2\pi \, \sigma^i_{3.6}\,\sigma^i_{4.5}} \times \exp\left({-\chi^2/2}\right) \; ;
\end{equation}
where:
\begin{equation}
\label{eq:chi2}
\chi^2= \left( \frac{ [3.6]^i_{obs} -[3.6]^j_{th}}{\sigma^i_{3.6}}\right)^2 + \left( \frac{ [4.5]^i_{obs} -[4.5]^j_{th}}{\sigma^i_{4.5}}\right)^2 \; ;
\end{equation}
here the j index runs on the masses along the isochrone and $\sigma^i$ are the
photometric uncertainties for the i-th star and subscripts \emph{obs} and
\emph{th} stand for observed and predicted quantities respectively. 
Since we use the magnitude-magnitude space the uncertainties can be considered
uncorrelated and the $\chi^2$ assumes the form of eq. (\ref{eq:chi2}). 
We assign to the i-th data point the mass value $m_{j*}$ for which
$\mathcal{L}(m_j)$ has its maximum. \\

%% \begin{figure}
%%  \centering
%%   \includegraphics[width=8cm]{match_20may/opt_nir_xexcirac.ps}
%%    \caption{{\it 1: }[[3.6]-[4.5], V] CMD and highlighted with different
%%      symbols are the stars in 2 consecutive horizotal slices with V between 20
%%      and 21 and with V between 21 and 22. {\it 2:} [[3.6]-[4.5],[3.6]] CMD
%%      with highlighted the same sources as in panel 1. The derived lines used to
%%      compute the disk fraction are overplotted.
%%             }
%%       \label{dfrmir}
%%  \end{figure}

 \begin{figure*}
  \centering
  \includegraphics[width=16cm]{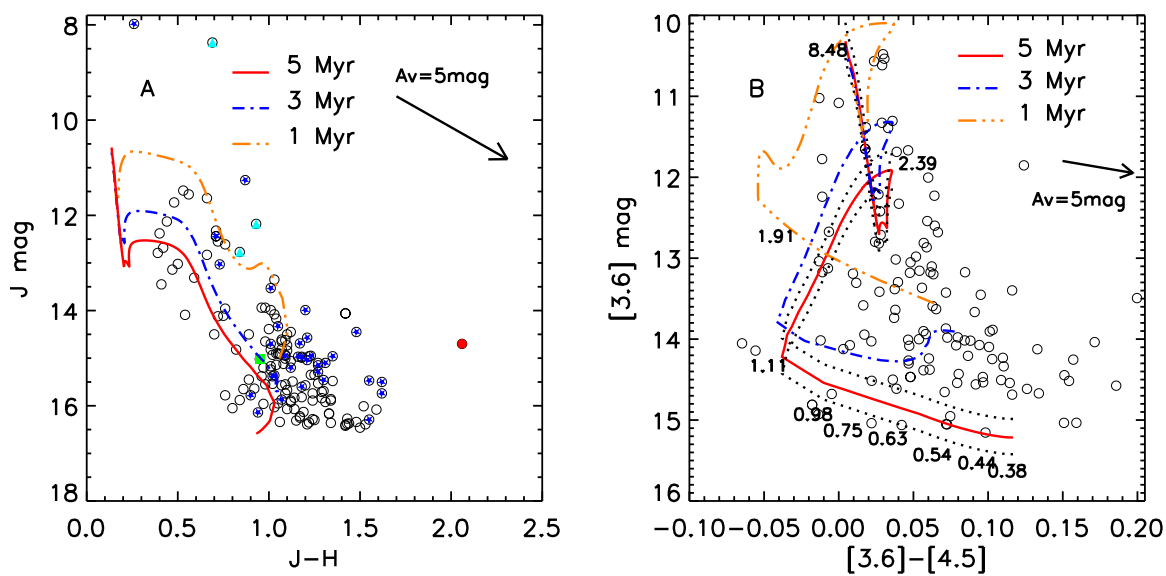}
  \caption{ {\it A:} CMD of the cluster members using the 2MASS magnitudes. Overlaid are the 1~Myr, 3~Myr and 5~Myr isochrones (shifted at a
    distance of the cluster of 2~kpc and Av = 2 mag). The different symbols and colors are like in Fig.~\ref{cc+am_all}. {\it B:} Close up of CMD in panel B in Fig.~\ref{cc+am_all} around sources
    without infrared excess (empty circles) in the IRAC CMD (panel B in
    Fig.~\ref{cc+am_all}).  
    Overlaid are the 1~Myr, 3~Myr and 5~Myr isochrones (shifted at a
    distance of the cluster of 2~kpc  and Av = 2 mag). 
    The dotted lines represent the 5~Myr isochrones but shifted at a maximum
    and minimum distance of 2.2 and 1.8~kpc. The numbers along the 5~Myr
    isochrone represent the corresponding masses (in $M_\odot$). 
    The arrow in panels {\it A} and {\it B} represents an extinction Av = 5 mag.
    }  
    \label{age}
 \end{figure*}


. Make sure there is at least one \tablenotemark
%% in the table for each \tablenotetext.
\tablecomments{Notation for the Infrared Excess:
  `\checkmark'=[3.6]-[4.5]$>$0.2mag; `-'=[3.6]-[4.5]$<$0.1mag;
  `$\sim$'=0.1$<$[3.6]-[4.5]$<$0.2mag.  
}
\tablenotetext{a}{Source position in IRAC/channel 2.}
\tablenotetext{b}{Source position in IRAC/channel 3.}
\tablenotetext{c}{The counterparts at different wavelengths were found using a
  different matching procedure.}
%  3.6'' and were original not included in the final IRAC photometry.}
\tablenotetext{d}{IRAC source with 2 X-ray counterparts (`X binary').}
\tablenotetext{e}{IRAC source with 3 X-ray counterparts.}
\end{deluxetable}

. Make sure there is at least one \tablenotemark
%% in the table for each \tablenotetext.
\tablecomments{The near-infrared fluxes are from the 2MASS catalog \cite{struskie2006},
  selecting only sources flagged 'AAA'. The optical fluxes are from Jos\'e et
  al. in preparation. The IRAC flux errors refer to the covariance matrix calculated from the
``best-fit" of the PRF to the data and do not include calibation
uncertainty ($\sim$10\% of the flux).  \\
}
%\tablenotetext{a}{The counterparts at different wavelengths have been found
%  using a different matching procedure.}
%\tablenotetext{b}{The IRAC counterparts have IRAC position errors larger than
%  3.6'' and were original not included in the final IRAC photometry.}
%\tablenotetext{c-m}{X-ray `binary'.}
\end{deluxetable}

\begin{deluxetable}{l rrrr l rrrr}
\tabletypesize{\scriptsize}
%\rotate
\tablecaption{Cluster member candidates: designation and positions of IRAC sources with excess without X-ray counterpart.\label{onlyirac-withexc-pos}}
\tablewidth{0pt}
\tablehead{
\colhead{ID} & \colhead{RA$_{\rm J2000}$} & \colhead{RA error} &
\colhead{DEC$_{\rm J2000}$} & \colhead{DEC error} 
\\
%}
%\tablehead{
\colhead{IRAC-} & \colhead{[deg]} & \colhead{$10^{-4}$[deg]} &
\colhead{[deg]} & \colhead{$10^{-4}$[deg]} 
}
\startdata
02270033+6205291&36.751377&1.09&62.091415&1.12\\
02271605+6200506&36.816875&0.43&62.014042&0.44\\
02271147+6201578&36.797775&0.57&62.032722&0.60\\
02271512+6200152&36.813011&0.45&62.004215&0.46\\
02271175+6158265&36.798965&0.65&61.974022&0.67\\
02270991+6158521&36.791275&1.63&61.981152&1.73\\
02264351+6205302&36.681278&3.98&62.091713&3.92\\
02264566+6204469&36.690266&0.60&62.079685&0.60\\
02270243+6200092&36.760136&1.02&62.002544&1.05\\
02263985+6205356&36.666031&5.19&62.093216&5.15\\
02264926+6203131&36.705231&1.63&62.053650&1.71\\
02265258+6202179&36.719074&2.50&62.038315&2.40\\
02271166+6157119&36.798573&1.27&61.953308&1.74\\
02265704+6200587&36.737675&2.37&62.016312&2.66\\
02265770+6200430&36.740398&0.95&62.011936&0.92\\
02264230+6204412&36.676254&2.29&62.078098&2.46\\
02263648+6206069&36.652004&1.45&62.101921&1.60\\
02265787+6200283&36.741112&1.13&62.007870&1.08\\
02264396+6204054&36.683167&5.15&62.068153&5.35\\
02270190+6159131&36.757904&7.53&61.986961&8.26\\
02264299+6204100&36.679127&4.42&62.069454&4.61\\
02270071+6159226&36.752964&0.38&61.989597&0.40\\
02264101+6204364&36.670876&2.35&62.076767&2.41\\
02270217+6158583&36.759029&2.45&61.982857&2.49\\
02265433+6200584&36.726395&1.50&62.016228&1.58\\
02265824+6159566&36.742676&2.14&61.999046&2.32\\
02270838+6157024&36.784935&1.46&61.950680&1.32\\
02265430+6200448&36.726261&2.01&62.012451&2.27\\
02265398+6200396&36.724911&0.50&62.010986&0.50\\
02270091+6158478&36.753777&0.83&61.979946&0.87\\
02265046+6201324&36.710236&1.51&62.025658&1.60\\
02263797+6204378&36.658195&6.49&62.077168&6.77\\
02270433+6157363&36.768024&1.67&61.960075&1.73\\
02264423+6202464&36.684288&0.89&62.046223&0.88\\
02265682+6159142&36.736732&2.36&61.987274&2.36\\
02264970+6200588&36.707077&0.91&62.016335&0.95\\
02265782+6158377&36.740898&2.64&61.977142&2.76\\
02264197+6202428&36.674873&1.04&62.045235&1.09\\
02264781+6201123&36.699203&0.92&62.020096&1.02\\
02262952+6205518&36.623009&0.86&62.097713&0.87\\
02264605+6201221&36.691887&0.59&62.022797&0.62\\
02264800+6200576&36.700001&1.96&62.016006&1.93\\
02270253+6157046&36.760521&1.12&61.951283&1.02\\
02264766+6200592&36.698593&1.67&62.016449&1.71\\
02265560+6158517&36.731667&1.41&61.981014&1.38\\
02264441+6201480&36.685043&1.40&62.030003&1.47\\
02264808+6200490&36.700348&1.44&62.013607&1.56\\
02264515+6201320&36.688110&2.02&62.025558&2.04\\
02264675+6201026&36.694805&1.56&62.017384&1.57\\
02262986+6205201&36.624416&2.08&62.088928&2.11\\
02265324+6159124&36.721840&1.76&61.986790&1.65\\
02265657+6158167&36.735699&1.53&61.971313&1.58\\
02264334+6201432&36.680584&1.64&62.028675&1.65\\
02265922+6157289&36.746742&2.32&61.958027&2.60\\
02263045+6204552&36.626877&1.53&62.081989&1.57\\
02265139+6159252&36.714127&1.92&61.990337&2.05\\
02264472+6201039&36.686325&1.61&62.017742&1.60\\
02264531+6200544&36.688805&1.36&62.015099&1.42\\
02264994+6159423&36.708065&0.64&61.995083&0.70\\
02263066+6204411&36.627762&6.30&62.078075&6.63\\
02264856+6159576&36.702320&2.02&61.999344&2.03\\
02265062+6159144&36.710911&2.52&61.987347&2.56\\
02264573+6200251&36.690525&0.75&62.006966&0.73\\
02262962+6204414&36.623398&1.58&62.078167&1.63\\
02264771+6159550&36.698799&2.01&61.998615&2.02\\
02263241+6203518&36.635040&3.08&62.064396&3.05\\
02265343+6158188&36.722610&2.31&61.971893&2.33\\
02264572+6200131&36.690487&1.06&62.003635&1.12\\
02262883+6204365&36.620129&0.95&62.076797&1.02\\
02264107+6201211&36.671124&1.73&62.022541&1.85\\
02263081+6203502&36.628380&1.64&62.063934&1.68\\
02263152+6203331&36.631325&1.96&62.059208&2.16\\
02263839+6201459&36.659939&0.76&62.029404&0.83\\
02264099+6201076&36.670776&2.30&62.018787&2.38\\
02264433+6200027&36.684719&0.83&62.000736&0.86\\
02264710+6159238&36.696262&0.96&61.989952&0.99\\
02263903+6201298&36.662609&1.50&62.024940&1.65\\
02264000+6200545&36.666649&0.65&62.015137&0.66\\
02263131+6203075&36.630451&3.48&62.052086&3.44\\
02265124+6157513&36.713505&1.87&61.964245&1.86\\
02265582+6156379&36.732563&1.11&61.943851&1.12\\
02265210+6157362&36.717083&1.41&61.960068&1.39\\
02263059+6203152&36.627441&3.69&62.054234&3.90\\
02265019+6157577&36.709145&1.87&61.966015&1.77\\
02264340+6159382&36.680817&0.73&61.993931&0.83\\
02263700+6201129&36.654163&6.04&62.020256&5.51\\
02264755+6158192&36.698109&0.95&61.972008&0.96\\
02263974+6200109&36.665581&0.72&62.003033&0.70\\
02263447+6201342&36.643616&0.79&62.026169&0.78\\
02263626+6200559&36.651093&1.18&62.015522&1.16\\
02264177+6159402&36.674049&1.28&61.994495&1.31\\
02262984+6202441&36.624340&1.77&62.045589&1.52\\
02262016+6205106&36.583988&2.51&62.086273&2.33\\
02264207+6159243&36.675278&2.06&61.990089&1.85\\
02262754+6203108&36.614742&3.30&62.053005&3.17\\
02262015+6205106&36.583969&2.90&62.086277&2.88\\
02263487+6201107&36.645279&1.39&62.019642&1.43\\
02263911+6200006&36.662975&1.08&62.000172&1.08\\
02263613+6200468&36.650528&1.92&62.013004&1.94\\
02264007+6159453&36.666950&1.66&61.995907&1.69\\
02262284+6204130&36.595158&2.46&62.070282&2.57\\
02264298+6158535&36.679089&1.58&61.981522&1.65\\
02262198+6204219&36.591591&1.54&62.072754&1.65\\
02263826+6159587&36.659424&1.31&61.999649&1.42\\
02263546+6200378&36.647755&1.99&62.010502&2.06\\
02264176+6158503&36.674004&0.62&61.980648&0.59\\
02262208+6204082&36.592018&4.33&62.068935&4.68\\
02263629+6200228&36.651199&3.34&62.006340&3.27\\
02261795+6204595&36.574806&0.43&62.083206&0.42\\
02263238+6201144&36.634914&2.44&62.020657&2.31\\
02263867+6159406&36.661110&1.95&61.994610&2.00\\
02264479+6157563&36.686630&2.25&61.965633&2.08\\
02264892+6156541&36.703846&1.88&61.948353&2.01\\
02263509+6200286&36.646210&1.78&62.007942&1.76\\
02262053+6203573&36.585541&0.56&62.065926&0.59\\
02262221+6203481&36.592525&1.07&62.063354&1.09\\
02262064+6204068&36.585991&0.75&62.068542&0.78\\
02262156+6203522&36.589840&0.92&62.064491&0.94\\
02262923+6201481&36.621777&0.40&62.030014&0.40\\
02263887+6159212&36.661957&1.25&61.989231&1.24\\
02262002+6203388&36.583420&0.27&62.060768&0.28\\
02264367+6157567&36.681976&1.02&61.965744&1.02\\
02264451+6157367&36.685448&2.66&61.960186&2.33\\
02264288+6157568&36.678658&1.90&61.965778&1.90\\
02263081+6201025&36.628357&1.92&62.017365&1.84\\
02263648+6159324&36.652008&2.23&61.992340&2.44\\
02263111+6200553&36.629620&1.99&62.015350&2.09\\
02262779+6201395&36.615810&1.60&62.027645&1.79\\
02263351+6200061&36.639633&1.80&62.001694&1.89\\
02264716+6156281&36.696507&2.23&61.941135&2.16\\
02262460+6202206&36.602512&1.52&62.039051&1.74\\
02264094+6157557&36.670597&1.35&61.965462&1.50\\
02264327+6157180&36.680298&1.04&61.955002&1.05\\
02264186+6157374&36.674408&2.20&61.960377&2.26\\
02263532+6159162&36.647175&1.53&61.987839&1.60\\
02263634+6159002&36.651421&3.04&61.983398&2.88\\
02261768+6203533&36.573673&3.44&62.064819&3.63\\
02264374+6157020&36.682247&1.75&61.950546&1.98\\
02261312+6204548&36.554668&3.02&62.081898&2.91\\
02263089+6200123&36.628719&2.08&62.003429&2.06\\
02263265+6159390&36.636040&4.43&61.994171&3.12\\
02263645+6158422&36.651855&2.04&61.978397&2.27\\
02262824+6200461&36.617680&1.31&62.012794&1.33\\
02263265+6159390&36.636051&2.22&61.994171&2.41\\
02263695+6158253&36.653969&0.89&61.973694&0.99\\
02262584+6201141&36.607647&1.97&62.020576&1.97\\
02263477+6158522&36.644855&1.29&61.981155&1.31\\
02263597+6158349&36.649887&2.12&61.976360&2.20\\
02264017+6157233&36.667385&0.66&61.956478&0.68\\
02263977+6157249&36.665695&0.66&61.956909&0.67\\
02262906+6200162&36.621078&1.39&62.004505&1.41\\
02262962+6200067&36.623421&1.56&62.001865&1.59\\
02263204+6159304&36.633484&2.83&61.991768&3.03\\
02261582+6203423&36.565907&7.35&62.061756&7.45\\
02263313+6158556&36.638035&0.72&61.982124&0.78\\
02262453+6201150&36.602200&1.52&62.020844&1.58\\
02263072+6159361&36.628014&1.18&61.993362&1.27\\
02263828+6157376&36.659508&1.02&61.960442&1.04\\
02262992+6159417&36.624687&5.48&61.994930&5.11\\
02262339+6201159&36.597473&6.28&62.021088&7.05\\
02263099+6159162&36.629112&3.54&61.987827&3.28\\
02262659+6200126&36.610809&1.13&62.003498&1.18\\
02262679+6200087&36.611645&1.28&62.002426&1.33\\
02260909+6204451&36.537888&6.82&62.079201&6.46\\
02262691+6200036&36.612110&1.90&62.001007&1.94\\
02263154+6158508&36.631401&2.36&61.980770&2.51\\
02261586+6202379&36.566078&6.78&62.043861&7.65\\
02262435+6200298&36.601448&2.38&62.008282&2.30\\
02261008+6203539&36.542019&5.68&62.064960&5.79\\
02261113+6203319&36.546360&3.74&62.058861&3.73\\
02263768+6156392&36.657013&1.31&61.944225&1.32\\
02261008+6203502&36.542004&6.24&62.063934&6.58\\
02263269+6157570&36.636223&1.69&61.965843&1.87\\
02261202+6203161&36.550064&1.05&62.054482&1.06\\
02262838+6159025&36.618248&3.24&61.984032&3.05\\
02261734+6201385&36.572254&6.22&62.027363&6.84\\
02262752+6159097&36.614651&3.06&61.986019&3.06\\
02260861+6204033&36.535892&3.52&62.067593&3.62\\
02261013+6203366&36.542225&1.40&62.060158&1.39\\
02261052+6203303&36.543831&0.94&62.058414&0.96\\
02263236+6157458&36.634819&2.66&61.962730&2.49\\
02262696+6159079&36.612316&0.95&61.985519&0.96\\
02262234+6200137&36.593098&0.97&62.003815&1.05\\
02263130+6157478&36.630428&3.25&61.963284&3.16\\
02262853+6158187&36.618858&0.90&61.971859&0.89\\
02260643+6204078&36.526806&4.97&62.068821&4.84\\
02260732+6203433&36.530510&2.05&62.062019&2.11\\
02263175+6157306&36.632278&2.73&61.958488&2.91\\
02262478+6159141&36.603256&0.66&61.987259&0.68\\
02261626+6201284&36.567768&4.48&62.024544&4.87\\
02260836+6203245&36.534851&0.50&62.056801&0.51\\
02260536+6204052&36.522316&0.76&62.068115&0.78\\
02260671+6203259&36.527969&3.31&62.057201&3.42\\
02262166+6159323&36.590260&1.66&61.992294&1.79\\
02260728+6203143&36.530346&5.05&62.053967&4.99\\
02260783+6203009&36.532642&0.61&62.050243&0.62\\
02261314+6201390&36.554729&4.11&62.027500&4.35\\
02262034+6159450&36.584766&1.00&61.995838&1.02\\
02260516+6203287&36.521511&3.63&62.057972&3.74\\
02262796+6157322&36.616486&1.15&61.958935&1.18\\
02261792+6200067&36.574650&2.00&62.001865&2.13\\
02260364+6203428&36.515186&6.42&62.061882&6.07\\
02262088+6159068&36.586994&2.30&61.985210&2.31\\
02261410+6200439&36.558743&2.35&62.012188&2.34\\
02260273+6203383&36.511391&5.38&62.060650&5.29\\
02260720+6202266&36.529999&7.25&62.040718&7.14\\
02261544+6200211&36.564346&1.99&62.005852&2.00\\
02260332+6203289&36.513847&4.63&62.058022&4.66\\
02260500+6202511&36.520821&0.50&62.047527&0.51\\
02260158+6203460&36.506577&1.89&62.062775&1.89\\
02260688+6202186&36.528656&5.35&62.038502&5.21\\
02261971+6158561&36.582123&1.08&61.982254&1.10\\
02260720+6202110&36.530014&5.89&62.036377&6.23\\
02260222+6203283&36.509258&7.22&62.057858&7.61\\
02260251+6203171&36.510445&1.62&62.054737&1.75\\
02261717+6159237&36.571560&0.84&61.989914&0.81\\
02261771+6159155&36.573807&0.67&61.987633&0.69\\
02262246+6157547&36.593582&1.46&61.965191&1.35\\
02260424+6202377&36.517670&5.61&62.043819&5.81\\
02260672+6201592&36.528019&3.75&62.033119&3.85\\
02260051+6203191&36.502117&9.31&62.055294&9.85\\
02260099+6203083&36.504139&1.70&62.052303&1.80\\
02260121+6202243&36.505024&0.41&62.040092&0.43\\
02261161+6159551&36.548382&0.98&61.998638&1.00\\
02262195+6157068&36.591454&2.66&61.951885&2.60\\
02260240+6201596&36.510014&5.35&62.033211&6.10\\
02255814+6203025&36.492268&6.66&62.050694&6.61\\
02260196+6202040&36.508171&3.50&62.034431&3.84\\
02261251+6159187&36.552143&2.83&61.988541&2.72\\
02261109+6159382&36.546188&2.54&61.993958&2.45\\
02255901+6201599&36.495857&3.49&62.033318&3.51\\
02255913+6202366&36.496380&2.55&62.043491&2.55\\
02260892+6159322&36.537155&1.05&61.992275&1.17\\
02255696+6202370&36.487343&7.45&62.043606&8.23\\
02261901+6156250&36.579205&2.06&61.940285&1.95\\
02255828+6201436&36.492817&4.57&62.028790&4.61\\
02255812+6201380&36.492149&7.19&62.027218&7.72\\
02260332+6200131&36.513851&1.17&62.003643&1.24\\
02260697+6159122&36.529022&2.11&61.986710&2.31\\
02261924+6155512&36.580158&1.03&61.930901&0.97\\
02260544+6158246&36.522675&0.35&61.973492&0.41\\
02260292+6158565&36.512161&0.60&61.982368&0.60\\
02260237+6158308&36.509884&0.49&61.975220&0.52\\
02260187+6158348&36.507786&0.64&61.976341&0.65\\
02260760+6156491&36.531658&1.15&61.946983&1.16\\
02260445+6157348&36.518532&4.54&61.959667&5.16\\
02255760+6158528&36.490013&4.85&61.981339&5.00\\
02262573+6203558&36.607193&0.31&62.065487&0.32\\
 \enddata
%% Text for table notes should follow after the \enddata but before
%% the \end{deluxetable}. Make sure there is at least one \tablenotemark
%% in the table for each \tablenotetext.
%\tablecomments{The near-infrared fluxes are from the 2MASS catalog (REF.),
%  selecting only sources flagged 'AAA'. The optical fluxes are from Jos\'e et
%  al. in preparation.}
%\tablenotetext{a}{The counterparts at different wavelengths have been found
%  using a different matching procedure.}
%\tablenotetext{b}{The IRAC counterparts have IRAC position errors larger than
%  3.6'' and were original not included in the final IRAC photometry.}
%\tablenotetext{c-m}{X-ray `binary'.}
\end{deluxetable}

. Make sure there is at least one \tablenotemark
%% in the table for each \tablenotetext.
\tablecomments{The IRAC flux errors refer to the covariance matrix calculated from the
``best-fit" of the PRF to the data and do not include calibation
uncertainty ($\sim$10\% of the flux).
}
%\tablenotetext{a}{The counterparts at different wavelengths have been found
%  using a different matching procedure.}
%\tablenotetext{b}{The IRAC counterparts have IRAC position errors larger than
%  3.6'' and were original not included in the final IRAC photometry.}
%\tablenotetext{c-m}{X-ray `binary'.}
\end{deluxetable}

\bibliographystyle{apj}
\bibliography{references.bib}

\end{document}